\documentclass[a4paper, 12pt]{article}
\usepackage{amsfonts,amsmath,amsthm}
\usepackage{graphicx}
\usepackage{authblk}
\usepackage{hyperref}
\usepackage{microtype}
\usepackage{lipsum}
\usepackage[section]{placeins}
\usepackage{afterpage}
\usepackage{capt-of}
\usepackage{pifont}
\usepackage{bm}
\usepackage{pdfpages}
\usepackage{mdframed}
\usepackage{multirow}
\usepackage{arydshln}

\newtheorem{thm}{Theorem}
\newtheorem{lem}[thm]{Lemma}

\newtheorem{dfn}[thm]{Definition}

\def\vol{\mbox{vol}}

\def\RR{{\mathbb R}}

\usepackage{geometry}
\begin{document}
\date{\quad}
\title{Modeling asset allocation strategies and a new portfolio performance score}
\author[1,2]{Apostolos Chalkis}
\author[1,2]{Emmanouil Christoforou}
\author[2,1]{Ioannis Z.~Emiris}
\author[2]{Theodore Dalamagas}

\affil[1]{Department of Informatics \& Telecommunications\linebreak National \& Kapodistrian University of Athens, Greece}
\affil[2]{ATHENA Research \& Innovation Center, Greece}

\maketitle

\begin{abstract}
We discuss and extend a powerful, geometric framework to represent the set of portfolios, which identifies the space of asset allocations with the points lying
in a convex polytope. Based on this viewpoint, we survey certain
state-of-the-art tools from geometric and statistical computing in order to
handle important and difficult problems in digital finance. Although our tools
are quite general, in this paper we focus on two specific questions.

The first concerns crisis detection, which is of prime interest for the
public in general and for policy makers in particular because of the significant
impact that crises have on the economy. Certain features in stock markets lead
to this type of anomaly detection: Given the assets’ returns, we describe the
relationship between portfolios’ return and volatility by means of a copula, without making any assumption on investor strategies. We examine a recent method relying on copulae to construct an
appropriate indicator that allows us to automate crisis detection. On real data, the indicator detects all past crashes in the cryptocurrency market, whereas from the DJ600-Europe index, from 1990 to 2008, the indicator identifies correctly~4 crises and issues one false positive for which we offer an explanation.

Our second contribution is to introduce an original computational framework to model asset allocation strategies, which is of independent interest for digital finance and its applications. Our approach addresses the crucial question
of evaluating portfolio management, and is relevant to individual managers as well as financial institutions. To evaluate portfolio performance, we provide a new portfolio
score, based on the aforementioned framework and concepts. In particular, our score relies on statistical properties of portfolios, and we show how they can be computed efficiently.
\end{abstract}

\newpage
\section{Introduction}

Modern finance has been pioneered by Markowitz who set a framework to study choice in portfolio allocation under uncertainty~\cite{M52}, and for which he earned the Nobel Prize in economics, 1990.
Within this framework, Markowitz characterized portfolios by their return and their risk; the latter is formally defined as the variance of the portfolios' returns\footnote{Throughout this paper we refer to the variance of a portfolio's return as {\em portfolio volatility}.}.
An investor builds a portfolio that would maximize its expected return for a chosen level of risk; it has since become common for asset managers to optimize their portfolio within this framework.
This approach has led a large part of the empirical finance research to focus on the so-called efficient frontier which is defined as the set of portfolios presenting the lowest risk for a given expected return. 
The efficient frontier is associated with a well-known family of convex functions, studied by Markowitz \cite{Markowitz56}. 
Moreover, the  distributional properties of the optimal portfolio weights have been used for efficient portfolio selection~\cite{Bodnar17,Bodnar16,Bodnar09,Raymond08,Jobson80}.

It is known, from the relevant literature, that financial markets exhibit three types of behavior. 
In normal times, stocks are characterized by slightly positive returns and moderate volatility, in up-market times (typically bubbles) by high returns and low volatility, and during financial crises by strongly negative returns and high volatility, see \cite{BGP12} for details.
So, following Markowitz' framework, in normal and up-market times, the stocks and portfolios with the lowest volatility should present the lowest returns, whereas during crises those with lowest volatility should present the highest returns. The detection of normal and crisis periods is also crucial for computing an efficient asset allocation~\cite{Ivanyuk21,Harzallah20,Pinho17}.

However, these tools, when used to build a portfolio, do not always guarantee a good performance in practice~\cite{MRT2010}. Thus, the analysis of investment performance is of special interest in modern finance, especially given the growth of the asset management industry in the last decades.
Research in this area is axed on Sharpe-like ratios proposed in the 1960’s \cite{Jensen67,Sharpe66,Treynor15}. In practice, the performance of a portfolio manager, over a given period, is usually measured as the ratio of his "excess" return, with respect to a benchmark portfolio, over a risk measure~\cite{Grinblatt94}. Managers are then ranked according to these ratios, and the one achieving the highest and steadiest returns receives the best score. The major drawback of these techniques is the identification of benchmark portfolios, since the formation of such portfolios remains controversial. 
Moreover, they suffer from non-negligible estimation errors \cite{Lo02}, which prevent any performance comparison to be significant. In \cite{P05} -and independently in \cite{Guegan11,Banerjee11}- they use a geometric representation of a stock market to define a cross-sectional score of a portfolio given a vector of assets' returns. The score of a portfolio is defined as the proportion of all possible asset allocations that the portfolio outperforms in terms of return. The aim is to measure the relative performance with respect to all possible alternative allocations offered to the manager. The term {\em cross section} is used to underline that the score takes into account portfolios that are diversified over all sections of assets, without studying -separately- the performance on specific sections of stocks. In \cite{Banerjee11}, they follow the same approach by defining what they call {\em naive investor's strategy}. A naive investor's strategy selects uniformly a portfolio from the set of portfolios, as it is agnostic of the asset returns generating process, and hence does not use any such information.

\subsection{Contributions}

First, we briefly survey the computational framework in~\cite{Cales18}, which uses the geometric representation of long-only portfolios in~\cite{P05} and the copula representation for the dependency between portfolios' return and volatility. A copula is a multivariate joint distribution where the marginal distributions are uniform; for more details on copulae, we refer to~\cite{Nelsen06}. We enhance this framework significantly by employing clustering methods on copulae, and we use it to detect all past crash events in the cryptocurrency market and all past crises from 1990 to 2008, using real data from DJ600.

We extend the geometric framework in~\cite{P05} to model additional asset allocations to long-only portfolios, e.g. the "150/50" or the "130/30" strategies, which recently have gained popularity~\cite{Patel07}. In particular, we work with the set of fully-invested portfolios, i.e., portfolios whose weights sum up to 1, which is the default choice for the bulk of the asset management industry. More precisely, we allow the weights be negative and we use the norm-constraint in~\cite{Zhao20} to set a lower bound on the weights' values. Then, we introduce a transformation to represent the set of all possible fully-invested portfolios by a convex polytope; i.e., each point in the interior of the polytope corresponds to a single asset allocation.

We use this geometric representation to introduce a new mathematical model of portfolio allocation strategies in a stock market. We consider the concept where portfolio managers compute and propose portfolio allocations, which we call {\em formal allocation proposals}. Then, an investor decides which asset allocation proposal to select. Second, she decides how much to modify this proposal to build her final portfolio. Thus, we expect the portfolios of the investors who have chosen the proposal of a manager to be "concentrated around" that proposal.
To model this procedure, we employ multivariate log-concave distributions. The support of the Probability Density Function (i.e.\ the subset of $\RR^n$ which is not mapped to zero) of each distribution is the set of all possible portfolios, i.e.\ a convex polytope. In particular, we say that a {\em portfolio allocation strategy} $F_{\pi}$ is induced from a log-concave distribution $\pi$ as follows: to build a portfolio with strategy $F_{\pi}$, sample a point/portfolio from $\pi$. Then, we call the mode of $\pi$ a {\em formal allocation proposal} of the allocation strategy $F_{\pi}$.

We use Markowitz's framework to parameterize the allocation strategies by the level of risk that a certain group of investors selects. Similarly, for a given level of risk, we use the variance to parameterize to what extent around the formal allocation proposal a subgroup of investors may decide to stick. 
Finally, as in any stock market, plenty of strategies may appear which are chosen by groups of investors. Thus, we define the {\em mixed strategy}, induced by a convex combination of log-concave distributions, i.e.\ a mixture distribution.

We use this model of portfolio allocation strategies to define a new portfolio score to evaluate the performance of an investment. Our new score considers the set of truly invested portfolios in a stock market in a given time period. 
We evaluate the performance of a portfolio, for a given time period, by comparing the portfolio against a mixed strategy $F_{\pi}$. Thus, we define the score of a portfolio as the expected number of truly invested portfolios that the first outperforms --in terms of return-- when the portfolios have been invested according to the mixed strategy $F_{\pi}$. To estimate the new cross-sectional score within an arbitrarily small error, we provide an efficient algorithm, based on Markov Chain Monte Carlo integration. In extreme cases, our new score becomes equal to that in \cite{P05,Guegan11,Banerjee11}. Thus, it can also be seen as a generalization of the latter cross-sectional score. 
Moreover, as one may have limited knowledge about how the investors behave in a stock market, or her knowledge may vary from a time period to another, we extend our framework to handle these issues. We also provide different versions of our score. Each version provides a piece of different information about the portfolio allocation we would like to evaluate.

We also provide an open-source implementation\footnote{\url{https://github.com/TolisChal/portfolio_scoring.git}} to simulate (mixed) allocation strategies and to compute our new score given a portfolio. Our implementation scales up to a few hundred assets and allocation strategies. We provide a pseudo-real time example in the cryptocurrency market, using the 12 cryptocurrencies with longest history. We provide extended arithmetic results to show that the informativeness of our new score can be higher than that of existing and well-known performance measures (e.g.\ Sharpe, Sortino ratios, and Jensen's alpha). Moreover, we use our computations of the distribution of a portfolio's score --assuming a distribution on the assets' returns-- to discuss how it could lead to useful insights about its performance. We also compute copulae of portfolios' return and volatility under the assumption that the portfolios have been built according to a mixed strategy. We show that a copula of a certain time period can be very different from that in~\cite{Cales18}. We believe that the last two simulations pave the way for future work in the problems of crisis detection and portfolio allocation.

Finally, since the simulation of allocation strategies and the computation of the score and copulae rely only on sampling from high dimensional log-concave distributions supported on the set of portfolios, our framework works also for a singular covariance matrix. That is, we can incorporate in our framework the results in~\cite{Gulliksson20,Mazur18,Mazur17,Mazur16,Pappas10}. However, to keep the presentation simple, in Sec.~\ref{sec:model_strategies} we assume that the covariance matrix of the asset returns is positive definite. More details about our computational method and its efficiency are found in Appendix~\ref{appnd:computational_methods}.

\paragraph{Paper structure.}
The next section presents our geometric representation of portfolios.
Sec.~\ref{sec:crises_detection} surveys our work on copulae and the ensuing crisis indicator; our approach is corroborated by two applications on real data.
Some elements of this section are presented in~\cite{Cales18}, but here we present a broader class of methods (i.e.\ clustering copulae) and a new result on the cryptocurrency market. 
Sec.~\ref{sec:model_strategies} introduces a new framework for modeling allocation strategies and evaluating portfolio performance by defining a new score of a portfolio. Sec.~\ref{sec:simulations} presents our pseudo-real time example on real data to illustrate our new framework and the usefulness of our new score. Finally, in Sec.~\ref{sec:future_work}, we briefly discuss conclusions and future work.

\section{Geometric representation of the set of portfolios}\label{sec:geometric_representation}

In this section we formalize the geometric representation of sets of portfolios with an arbitrary large number of assets $n$. First we handle the case of long-only strategies and then, we extend this representation to fully-invested portfolios. In both cases the set of portfolios is a convex polytope in $\RR^n$.

\subsection{Long-only portfolios}

In this case no short sales are allowed.
Let a portfolio $x$ investing in $n$ assets, whose weights are $x=(x_1, \dots, x_n)\in\RR^{n}$. 
The portfolios in which a long-only asset manager can invest are subject to $\sum\limits_{i=1}^{n} x_i = 1$ and $x_i\geq 0, \forall i$. 
Thus, the set of portfolios available to this asset manager is the unit $(n-1)-$dimensional canonical simplex, denoted by $\Delta^{n-1}$ and defined as
\begin{equation}\label{eq:1}
\Delta^{n-1} := \left\{ x \in \mathbb{R}^{n}\ \left|\ \sum_{i=1}^{n} x_i=1, \mbox{ and } x_i \ge 0, i \in [n] \right.\right\}\subset \RR^n .
\end{equation}

The simplex $\Delta^{n-1}$ is the smallest convex polytope with nonzero volume in a given dimension.
For instance, in the plane any triangle is a simplex, while a triangular pyramid, or tetrahedron, is the simplex in 3d space. 
%
%
The vertices of $\Delta^{n-1}$ represent portfolios composed entirely of a single asset.

\subsection{Fully-invested portfolios}

When short sales are allowed we write the set of all possible portfolios as,

\begin{equation}
   P := \left\{ x \in \mathbb{R}^{n}\ \left|\ \sum_{i=1}^{n} x_i=1, \mbox{ and } \|x\|_1 \leq \gamma, i \in [n],\ \gamma \geq 1 \right.\right\}\subset \RR^n ,
\end{equation}
where the $L_1$-norm $\| x \|_1 = \sum_{i=1}^n |x_i|$. When $\gamma = 1$ no short sales are allowed and $P = \Delta^n$. When $\gamma = 1.6$ then $P$ corresponds to fully-invested portfolios of the 130/30 type and $\gamma = 2$ to 150/50. To show that $P$ is a convex polytope for any $\gamma\geq 1$, we replace the norm-constraint $\| x \|_1 \leq \gamma$ with a set of linear inequalities. Since $|x_i| = \max\{-x_i, x_i\}$, for each $x_i$, we add an auxiliary variable $y_i$ such that,
\begin{equation}
y_i \geq -x_i,\ y_i \geq x_i,\ y_i \geq 0 .
\end{equation}
Then, the set of all possible portfolios is given by,

\begin{equation}
\small
    \tilde{P} := \left \{ (x,y)\in\RR^{2n}\ \bigg|\ \sum_{i=1}^nx_i=1,\ -y_i \leq x_i \leq y_i,\ \sum_{i=1}^n y_i \leq\gamma,\ i \in [n],\ \gamma \geq 1 \right \}\subset \RR^{2n} ,
\end{equation}
which is a convex polytope as the feasible space is defined only by a set of linear inequalities (half-spaces). 

\section{Crisis detection}\label{sec:crises_detection}

In this section, we present our computational methods to address the problem of crisis detection in stock markets. We focus on long-only portfolios, which means that the set of portfolios in the following computations corresponds to the canonical simplex $\Delta^{n-1}$. 

\begin{figure}[t]
	\centering
\includegraphics[width=0.6\textwidth]{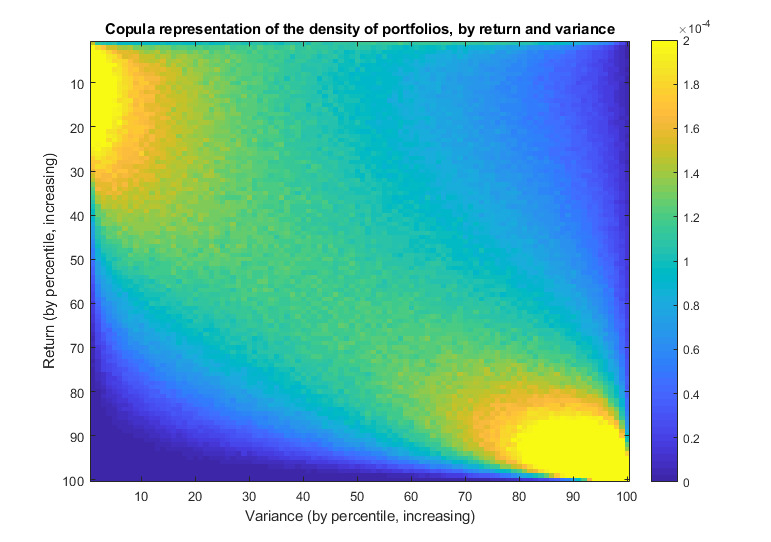}
	\caption{Copula representation of the portfolios distribution, by return and variance. The market considered is made of the 19 sectoral indices of DJSTOXX 600 Europe. The data is from Oct.~16, 2017 to Jan.~10, 2018. Each line and column sum to 1\% of the portfolios. 
	\label{fig:Illustration_Copula}}
\end{figure}

It is difficult to capture the dependency between portfolios return and volatility by the usual mean-variance representation. So we rely on the copula representation. 
A copula is a joint probability distribution for which all the marginal probability distributions are uniform. 
Fig.~\ref{fig:Illustration_Copula} illustrates such a copula and shows a positive dependency between portfolios' return and volatility.
Given a vector of assets' returns $R\in\RR^n$ and the covariance matrix $\Sigma\in\RR^{n\times n}$ of the assets' returns distribution, we say that any portfolio $x \in \Delta^{n-1}$ has return $f_{ret}(x, R) = R^T x$ and variance (or volatility) $f_{vol}(x, \Sigma) =  x^T\Sigma x$.

To estimate the copula between portfolios' return and volatility we consider the following discretization on the values of each quantity. We fix two sequences $s_0<\dots <s_m$ and $u_0<\dots <u_m$ such that 
\begin{equation}\label{eq:fracs}
\frac{\vol(S_i)}{\vol(\Delta^{n-1})} \approx p\quad\text{ and }\quad\frac{\vol(U_i)}{\vol(\Delta^{n-1})} \approx p,\; i=0,\dots ,m-1,
\end{equation}
where $S_i:=\{ x\in\RR^n\ |\ s_i\leq f_{ret}(x, R)\leq s_{i+1}\}$ and $U_i:=\{ x\in\RR^n\ |\ u_i\leq f_{vol}(x, \Sigma)\leq u_{i+1}\}$ and $p<1$ a small constant (e.g.\ $p=0.01$). Equation~(\ref{eq:fracs}) implies that a constant percentage $p$ of the portfolios have return less than $s_{i+1}$ and higher than $s_i$. The same occurs for all the sets $U_i$, which contain portfolios with bounded volatility. 

Furthermore, the sets $S_i,\ U_i$ define a grid of convex bodies, obtained by a family of parallel hyperplanes and a family of concentric ellipsoids --centered at the origin-- intersecting $\Delta^{n-1}$. In fact, for given integers $i,j\leq m-1$ the body
\begin{equation}
    Q_{ij} :=\{ x\in\Delta^{n-1}\ |\ s_i\leq f_{ret}(x, R)\leq s_{i+1} \text{ and } u_j\leq f_{vol}(x, \Sigma)\leq u_{j+1}\} ,
\end{equation}
contains the portfolios with return less than $s_{i+1}$ and higher than $s_i$ and volatility less than $u_{j+1}$ and higher than $u_j$. Now, to obtain the aforementioned copula one has to estimate the ratios $\frac{\vol(Q_{ij})}{\vol(\Delta^{n-1})}$ for $i,j=0,\dots ,m-1$. 

We use Monte Carlo to estimate each volume ratio. We leverage direct, efficient uniform sampling from $\Delta^{n-1}$ following \cite{RbMel98} and then count the number of points per body in the grid. In Sec.~\ref{subsec:indicator} this leads to an indicator to decide the state of the stock market that the estimated copula corresponds to. 

Considering the computational efficiency of this method, it can be applied to stock markets with a few thousands of assets, since the cost per uniformly distributed sample in $\Delta^{n-1}$ using the exact sampler in~\cite{RbMel98} is $O(n)$. For runtimes see Appendix~\ref{appnd:computational_methods}. 



\subsection{Computing copulae}\label{sec:computing_copulas}

In our computations, to define the family of parallel hyperplanes, we consider compound returns over periods of $k$ observations.
Let the asset returns $r_i=(r_{i,1},\dots,r_{i,n}) \in \mathbb{R}^{n}$, $i\in [k]$, then the component $j$ of the compound return equals:
\begin{equation}
    R_j = (1+r_{i,j})(1+r_{i+1,j})\cdots (1+r_{i+k-1,j}) -1, \quad j=1,\dots, n.
\end{equation}
This defines vector $R\in\RR^n$ normal to a family of parallel hyperplanes, whose equations are fully defined by selecting appropriate constants. 

The covariance matrix $\Sigma$ of the assets' returns is computed using the shrinkage estimator of \cite{LW04},\footnote{Matlab code at \scriptsize\url{http://www.econ.uzh.ch/en/people/faculty/wolf/publications.html}.} as it provides a robust estimate even when the sample size is short with respect to the number of assets.

To compute the copulae, we determine constants defining hyperplanes and ellipsoids so that the volume between two consecutive such objects is $p = 1\%$ of the simplex volume. 
Let us refer to the method outlined at Equation~(\ref{eq:fracs}) using notation introduced just before this equation.
The sequence of $s_0<\dots <s_m$ are determined by bisection using Varsi's algorithm.
For ellipsoids, we sample from the simplex and look for $u_0<\dots <u_m$ such that there is an equal number of uniformly distributed points in each intersection.  

\begin{figure}[t]
\includegraphics[width=0.33\textwidth]{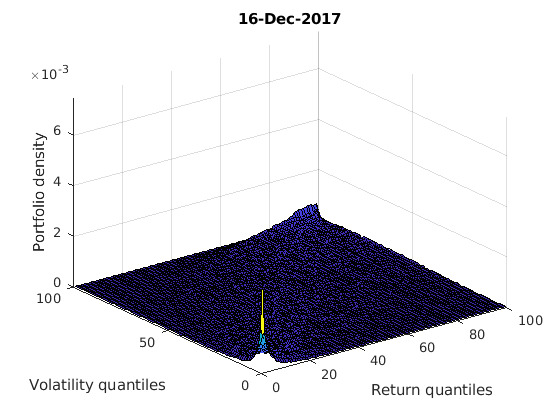} 
\includegraphics[width=0.3\textwidth]{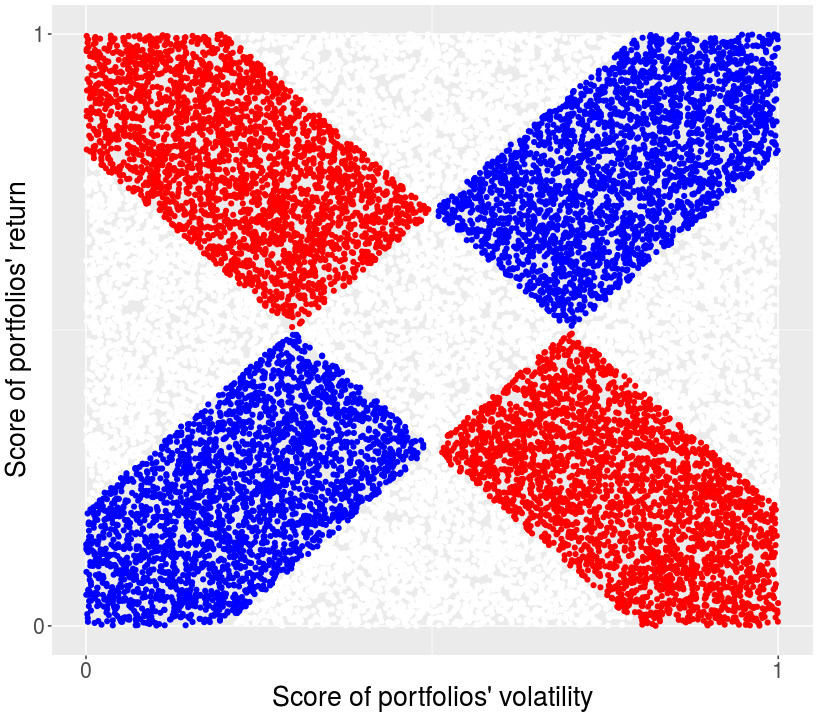}
\includegraphics[width=0.33\textwidth]{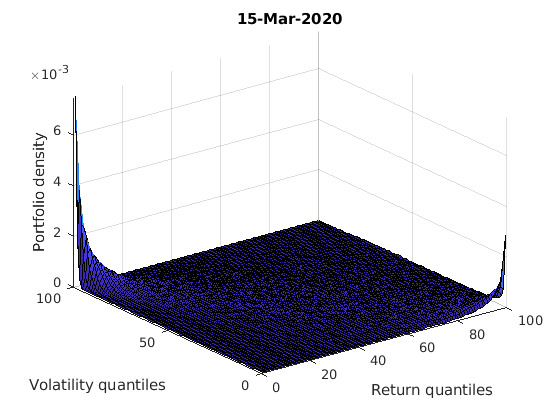} 
	\caption{Copulae that correspond to cryptocurrencies' states. Left, a normal period (16/12/2017) and right, a shock event due to Covid-19 (15/03/2020). The middle plot shows the mass of interest to characterize the market state.
	\label{fig:Returns_variance_relationship}}
\end{figure}

We set $m=100$, to estimate each copulae. 
We thus get $100\times 100$ copulae representing the distribution of the portfolios with respect to the portfolio returns and volatilities. 
Fig.~\ref{fig:Returns_variance_relationship} illustrates such copulae, and shows the different relationship between returns and volatility in good (left) and bad (right, Covid-19 shock event) times.

We analyze real data consisting of regular interval (e.g.\ daily) returns from two different asset sections: stocks from the Dow Jones Stoxx 600 Europe\texttrademark (DJ600) and cryptocurrencies. 
We apply the methodology to a subset of assets drawn from the DJ~600 constituents using daily data covering the period from 01/01/1990 to 31/11/2017\footnote{Our data is from Bloomberg\texttrademark.}.
Since not all stocks are tracked for the full period of time, we select the 100 assets with the longest history in the index, and juxtapose stock returns and stock returns covariance matrix over the same period to detect crises. 
For the cryptocurrency assets, we use the daily returns of 12 out of the top 100 cryptocurrencies, ranked by CoinMarketCap's\footnote{\scriptsize\url{https://coinmarketcap.com/}} market cap (cmc\_rank) on 22/11/2020, having the longest available history (Table~\ref{Tab:cryptos}). We compute the daily return for each coin using the daily close price obtained by CoinMarketCap, for several notable coins such as Bitcoin, Litecoin and Ethereum.

\begin{figure}[t]
		\includegraphics[width=1.0\textwidth]{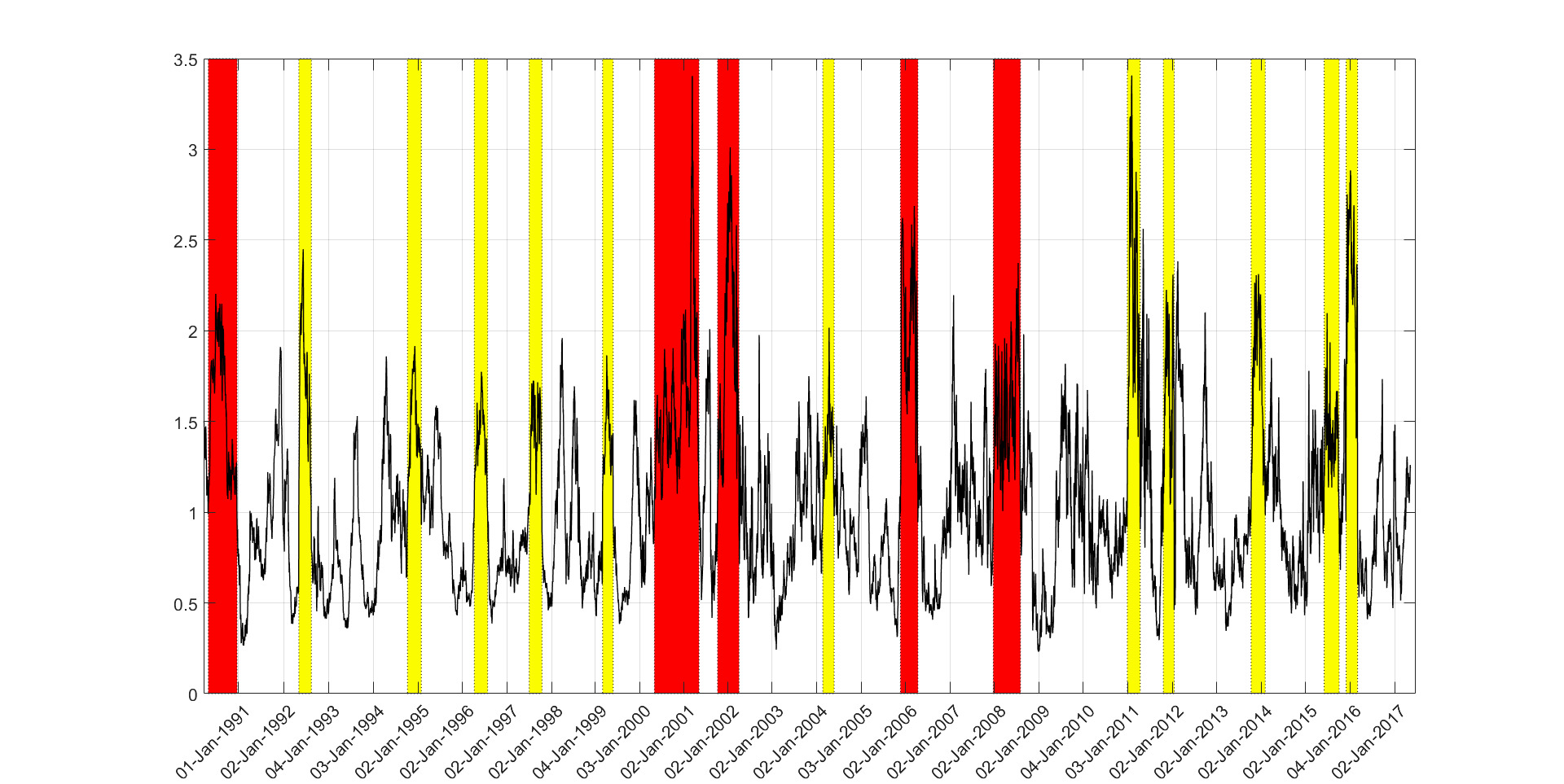}
	\caption{Representation of the periods over which the indicator is greater than one for 61-100 days (yellow) and over 100 days (red).
	\label{fig:WarningsCrises}}
\end{figure}

\subsection{Indicator and crisis detection}\label{subsec:indicator}

When we work with real data in order to build the indicator, we wish to compare the densities of portfolios along the two diagonals. In normal and up-market times, the portfolios with the lowest volatility present the lowest returns and the mass of portfolios should be on the up-diagonal. During crises, the portfolios with the lowest volatility present the highest returns and the mass of portfolios should be on the down-diagonal, see  Fig.~\ref{fig:Returns_variance_relationship} as illustration.
Thus, setting up- and down-diagonal bands, we define the indicator as the ratio of the down-diagonal band over the up-diagonal band, discarding the intersection of the two. The construction of the indicator is illustrated in Fig.~\ref{fig:Returns_variance_relationship} (middle) where the indicator is the ratio of the mass of portfolios in the blue area over the mass of portfolios in the red one.

\begin{figure}[t]
    \centering
    \includegraphics[width=1.0\linewidth]{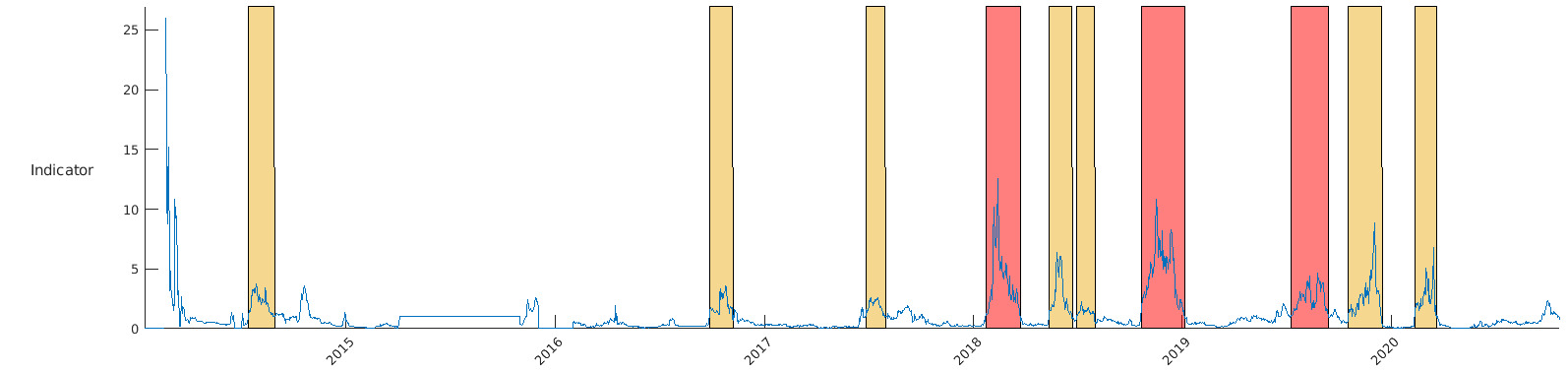}
    \caption{Warning (yellow) and Crises (red) periods detected by the indicator for cryptocurrencies (2014-2020).}
    \label{fig:crypto_indicator}
\end{figure}

The indicator is estimated on copulae by drawing $500,000$ uniformly distributed points.
We compute the indicator per copula over a rolling window of $k=60$ days and with a band of $\pm 10\%$ with respect to the diagonal. We experimentally determined both values. The latter corresponds to roughly 3 months when observations are daily. When the indicator exceeds 1 for more than 60 days but less than 100 days, we report the time interval as a ``warning'' (yellow color), while when exceeds~1 for more than 100 days, we report the interval as a ``crisis'' (red); see Fig.~\ref{fig:WarningsCrises}, and Fig.~\ref{fig:crypto_indicator}. 
The periods are at least 60 days long to avoid detection of isolated events whose persistence is only due to the auto-correlation implied by the rolling window. 

We compare DJ~600 results with the database of financial crises in European countries in~\cite{ESRB17}. The first crisis (May 1990 to Dec.\ 1990) corresponds to the early 90's recession, the second one (May 2000 to May 2001) to the dot-com bubble burst, the third one (Oct.~2001 to Apr.~2002) to the stock market downturn of 2002, the fourth one (Nov.~2005 to Apr.~2006) is not listed in the European database and is either a false positive of our method or may be due to a bias in the companies selected in the sample, and the fifth one (Dec.~2007 to Aug.\ 2008) can be associated with the sub-prime crisis.

Our cryptocurrencies indicator detects successfully the 2018 (great) cryptocurrency crash; see Fig.~\ref{fig:crypto_indicator}. The first shock event detected in 2018 (mid-January to late March) corresponds to the crash of nearly all cryptocurrencies, following Bitcoin's, whose price fell by about 65\% from 6 January to 6 February 2018, after an unprecedented boom in 2017. 
Intermediate warnings (mid-May to early August) should correspond to cryptocurrencies collapses (80\% from their peak in January) until September. 
The detected crash at the end of 2018 (November 2018 until early January 2019) corresponds to the fall of Bitcoin's market capitalization (below \$100 billion) and price by over 80\% from its peak, almost one-third of its previous week value. Finally, the detected event in early 2020 corresponds to the shock event due to covid-19.

\subsubsection{Clustering of copulae agrees with indicator}

To cluster the probability distributions distances of the copulae, we computed a distance matrix ($D$) between all copulae using the earth mover's distance (EMD) \cite{yossi2000}. 
The EMD between two distributions is the minimum amount of work required to turn one distribution into the other. We use a fast and robust EMD algorithm, which appears to improve both accuracy and speed~\cite{pele2009}. Then, we apply spectral clustering \cite{Andrew2001}, a method to cluster points using the eigenvectors of the affinity matrix ($A$) which we derive from the distance matrix, computed by the radial basis function kernel, replacing the Euclidean distance with EMD, where $A_{ij}=exp(-D_{ij}^2/2\sigma^2)$, and for $\sigma$ we chose the standard deviation of distances.
Using the $k$ largest eigenvectors of the laplacian matrix, we construct a new matrix and apply k-medoids clustering by treating each row as a point, so as to obtain $k$ clusters. The results with $k=6$ and $k=8$ are shown on the indicators' values in Fig.~\ref{fig:copulaeClusteringC6_clusters}, \ref{fig:copulaeClusteringC6EMD}, and \ref{fig:copulaeClusteringC8EMD}. Clusters appear to contain copulae with similar indicator values. Crisis and normal periods are assigned to clusters with high and low indicator values respectively. Therefore, the clustering of the copulae is proportional to discretising the values of the indicator.
We do not use any data-driven techniques to select an optimal cluster size, since we apply clustering only to demonstrate that the resulting clusters validate the indicator and distinguish different market states according to the indicator. Additional results on clustering copulae can be found in Appendix~\ref{appnd:clustering_copulae}.

\begin{figure}
    \centering
    \includegraphics[width=\textwidth]{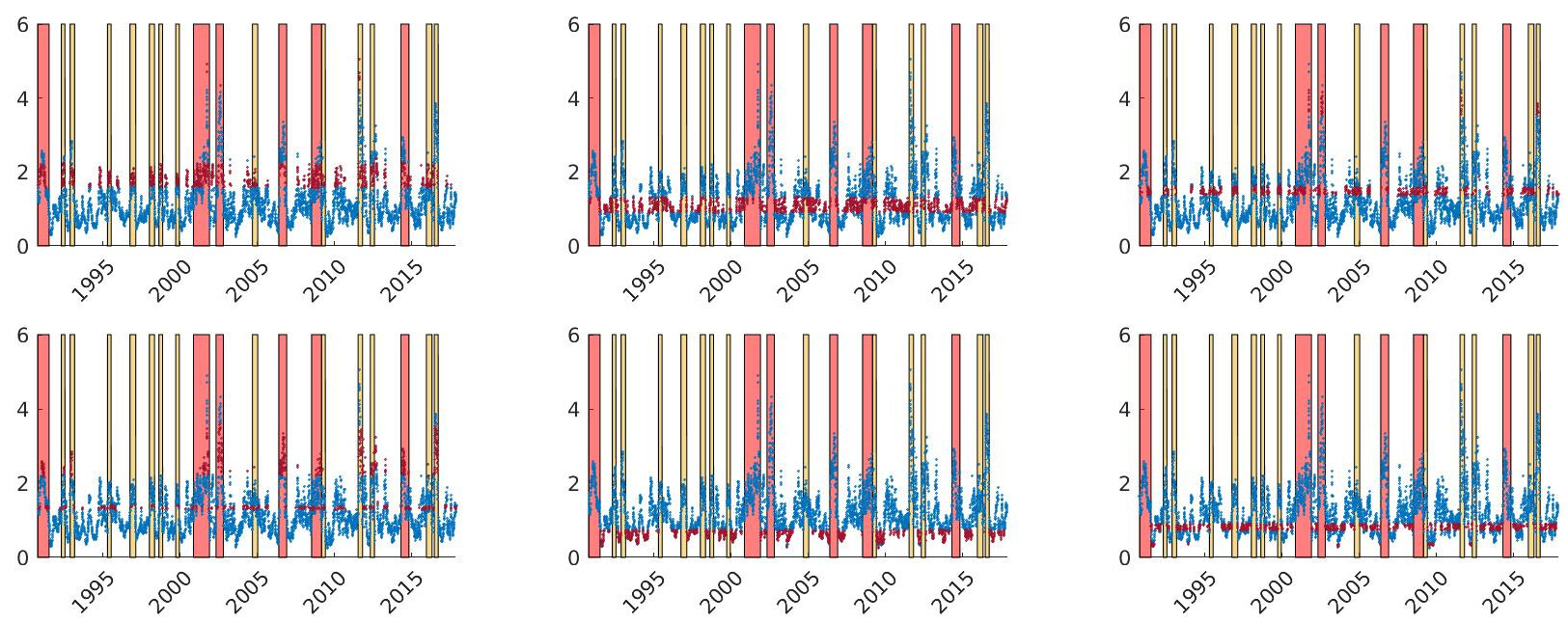}
    \caption{Spectral clustering of copulae, with $k=6$ clusters, on the earth mover's distances (EMD) of the copulae. Results are shown on the values of the indicator for every copula. There are 6 different plots, one for every cluster. Red points indicate the copulae assigned to the specific cluster, while the blue points are the copulae assigned to other clusters. Yellow and red time intervals are the identified by the indicator warning and crises periods respectively.}
    \label{fig:copula_clustering_emd_spectral}
\end{figure}


\section{Modeling allocation strategies and a new portfolio score}\label{sec:model_strategies}

We provide an original framework for modeling allocation strategies and a new cross-sectional portfolio score. We define the score of a given asset allocation as the expected value of the proportion of truly invested portfolios in a stock market, that the first outperforms when the portfolios have been built according to, what we call, a {\em mixed strategy}. 

Here, we assume that in a stock market the portfolio managers make allocation proposals. Then, the investors choose which proposal to select and how much to modify it before they build their final portfolio. Thus, we model a portfolio allocation strategy by a log-concave distribution supported on the portfolio domain $P$, with its mode being at a benchmark portfolio. Then, an investor builds a portfolio according to that strategy, by generating a point/portfolio from the corresponding distribution.

\begin{dfn}\label{def:strategy}
Let $\pi$ be a log-concave distribution supported on the portfolio domain $P\subset\RR^n$ with Probability Density Function (PDF) $\pi(x)$. Then, a portfolio allocation strategy $F:\pi\rightarrow P$ is said to be induced by the distribution $\pi$, and we write $F_{\pi}$.  
More precisely, $F_{\pi}$ is induced by the following state: 
\begin{center}
    \textit{``To build a portfolio with strategy $F_{\pi}$ sample a point/portfolio from $\pi$"}.
\end{center}
\end{dfn}

The mode of $\pi$ can be seen as the allocation proposal that a portfolio manager has made. Then, we expect the portfolios of the investors, who have chosen that proposal, to be concentrated around that proposal/mode.
\begin{dfn}
Let strategy $F_{\pi}$ induced by the unimodal distribution $\pi$.
We call the mode of $\pi$ formal allocation proposal or formal proposal of the portfolio allocation strategy $F_{\pi}$.
\end{dfn}

In the sequel, we assume that in a stock market the set of truly invested portfolios, are being built by a combination of different strategies used by the investors (mixed strategy). 
First, we consider a sequence of log-concave distributions $\pi_1,\dots ,\pi_M$ restricted to $P$. Each distribution induces a portfolio allocation strategy, i.e.\ $F_{\pi_1},\dots ,F_{\pi_M}$. Then, the mixed strategy is induced by a convex combination of $\pi_i$, i.e.\ by a mixture distribution, as the following definition states.
\begin{dfn}\label{def:mixed_strategy}
Let $\pi_1, \dots ,\pi_M$ be a sequence of log-concave distributions supported on the set of portfolios $P\subset\RR^n$, and let the mixture density be $\pi(x) = \sum_{i=1}^Mw_i\pi_i(x)$, where $w_i\geq 0,\ \sum_{i=1}^Mw_i=1$. We call $F_{\pi}$ the mixed strategy induced by the mixture density $\pi$.
\end{dfn}
\noindent 
In Definition~\ref{def:mixed_strategy} each weight $w_i$ corresponds to the proportion of the investors that build their portfolios according to the allocation strategy $F_{\pi_i}$. Thus, the vector of weights $w\in\RR^M$ implies how the investors, in a certain stock market and time period, tend to behave. 
Now we are ready to define the new cross-sectional score of an asset allocation versus a mixed strategy.

\begin{dfn}\label{def:score}
Let a stock market with $n$ assets and $F_{\pi}$ a mixed strategy induced by the mixture density $\pi$. For given asset returns $R\in\RR^n$ over a single period of time, the score of a portfolio, providing a value of return $R^*$, is
\begin{equation}\label{eq:new_score}
s = \int_{P} g(x)\pi (x) dx,\quad g(x) = \left\{
\begin{array}{ll}
      1. & \mbox{ if } R^Tx\leq R^* ,\\
      0, & \mbox{otherwise.}\\
\end{array} 
\right. 
\end{equation}
\end{dfn}
Clearly, the value of the integral in Equation~(\ref{eq:new_score}) corresponds to the expected proportion of portfolios that an allocation outperforms --in terms of return-- when the portfolios are invested according to the mixed strategy $F_{\pi}$.

\subsection{Log-concave distributions in Markowitz' framework} \label{subsec:markowitz_framework}

In this section, we model allocation strategies in Markowitz's framework using special multivariate log-concave distributions supported on the set of portfolios $P$. A proper choice of log-concave distributions allows us to parameterize a strategy by the level of risk and the level of dispersion around the formal allocation proposal of the strategy. 

In general, using Markowitz' framework one can define, under certain assumptions, the optimal portfolio $\bar{x}$ as the maximum of a concave function $h(x),\ x\in P$. Then, the mode of the log-concave distribution with PDF $\pi(x)\propto e^{\alpha h(x)}$ is $\bar{x}$, while the parameter $\alpha > 0$ controls the variance of the distribution. Large/small values of $\alpha$ corresponds to small/large variance.

Notice that as the variance grows, $\pi$ converges to the uniform distribution. Moreover, as the variance diminishes, the mass of $\pi$ concentrates around the mode of $\pi(x)$. Consequently, we use the variance to parameterize the sequence $\pi_i\propto e^{\alpha_ih(x)}$. That is, small variances correspond to allocation strategies used by investors who stick around the formal allocation proposal. Large variances correspond to allocation strategies used by investors who may modify the formal proposal a lot. Thus, in the first case, the invested portfolios would be highly concentrated around the formal allocation proposal of $F_{\pi}$ (or around the mode of $\pi$). In the second case, the invested portfolios would be highly dispersed around the mode of $\pi$. 

In the extreme case of very large variance, $\pi$ is close to the uniform distribution. Then, the induced allocation strategy becomes the naive strategy as defined in~\cite{Banerjee11}. 
We employ the $L_2$ norm of a log-concave distribution $\pi$ with respect to (w.r.t.)  the uniform distribution to characterize how dispersed, around the formal proposal, the portfolios built according to $F_{\pi}$ are. The $L_2$ norm of a distribution $f$ w.r.t\ a distribution $g$, when both are supported on a set $P\subset\RR^n$ is,

\begin{equation}
    \| f/g \| = \mathbb{E}_f\bigg( \frac{f(x)}{g(x)} \bigg) = \int_P \frac{f(x)}{g(x)}f(x)dx = \int_P\bigg( \frac{f(x)}{g(x)} \bigg)^2 g(x)dx .
\end{equation}
We can now define what we call a $D$-dispersed allocation strategy.

\begin{dfn}\label{def:dis_unif}
Let $\pi\propto e^{\alpha h(x)}$ be any log-concave distribution supported on the set of portfolios $P$ and let $F_{\pi}$ be the induced portfolio allocation strategy. We say that $F_{\pi}$ is $D$-dispersed, where $D$ is the $L_2$ norm of $\pi$ w.r.t.\ the uniform distribution.
\end{dfn}

\begin{figure}[t] \centering
\includegraphics[width=1.03\linewidth]{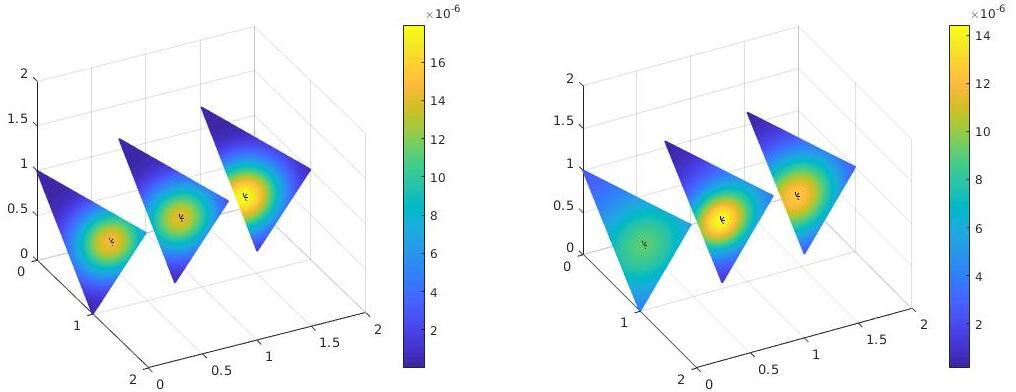}
\caption{Left: illustration of PDFs $\pi_q\propto e^{-\alpha \phi_q(x)}$, where $\alpha =1$ and from left to right $q_1 = 0.3,\ q_2 = 1,\ q_3 = 1.5$.  Right: $3$ illustrations of the mixture density of Equation~(\ref{eq:mix_density2}), where $M_1=3,\ M_2=2$. In both plots each black small star corresponds to a formal allocation proposal of an allocation strategy. From yellow to blue: high to low density regions. \label{fig:density}}
\end{figure}

Our main approach is to leverage the expected quadratic utility function,
\begin{equation}\label{eq:mean_variance_function}
    \phi_q(x) = x^T\Sigma x - q\mu^Tx,\ x\in P\subset\RR^n,\quad q\in [0,+\infty],
\end{equation}
where $\mu\in\RR^n$ is the mean and $\Sigma\in\RR^{n\times n}$ is the covariance matrix of the assets' returns and $n$ is the number of assets. 
This parametric function delivers similar solutions to the original Markowitz problem in~\cite{Kroll84,Levy79}. It is also used by the investors to compute the efficient frontier and optimal portfolios. The $x^T\Sigma x$ is called {\em risk term}, the $\mu^Tx$ is called {\em return term} and the parameter $q$ controls the trade-off between return and risk. Typically, in modern finance, a portfolio manager builds an efficient asset allocation by selecting a value $q_0$ ---which determines the level of risk of his allocation. Then, according to~\cite{Markowitz56} she/he solves the following optimization problem:  
$$
\min\; \phi_{q_0}(x) = x^T\Sigma x - q_0\mu^Tx,\;\text{ subject to } x\in P.
$$
We call the portfolio $\bar{x}=\min\limits_{x\in P}\phi_{q_0}(x)$ as the {\em optimal mean-variance} portfolio for the risk implied by $q_0$.
Thus, the efficient frontier can be seen as a parametric curve on $q$.

Let the log-concave distribution,
\begin{equation}
    \pi_{\alpha, q}\propto e^{-\alpha \phi_q(x)} ,
\end{equation}
supported on $P$. 
The left plot in Fig.~\ref{fig:density} illustrates some examples of the probability density function $\pi_{\alpha, q}$ where the mean $\mu$ and the covariance matrix $\Sigma$ are randomly sampled once. Notice that for different $q$, the mode (or the formal allocation proposal of the strategy $F_{\pi_{\alpha,q}}$) is shifted.

We use the parameter $q$ to denote the level of risk of a portfolio allocation strategy $F_{\pi_{\alpha,q}}$. Small values of $q$ correspond to low risk strategies whereas large values of $q$ to high risk strategies. Thus, a sequence of such densities can be parameterized by both $q$ (risk) and $\alpha$ (dispersion). In particular, a mixed strategy $F_{\pi}$ can be induced by the following mixture density:
\begin{equation}\label{eq:mix_density2}
\pi(x) = \sum_{i=1}^{M_1}\sum_{j=1}^{M_2} w_{ij}e^{-a_{ij}\phi_i(x)},\ \text{where }\phi_i = x^T\Sigma x - q_i\mu^Tx ,\ x\in P ,
\end{equation}
where each $q_i$ denotes the level of risk. For each $q_i$ the parameters $\alpha_{ij}$ imply the level of dispersion of the strategy $F_{\pi_{ij}}$. Notice that for each level of risk $q_i$ there are $M_2$ different levels of dispersion that different groups of investors' portfolios may appear around the same formal allocation proposal.
The right plot of Fig.~\ref{fig:density} illustrates some examples of this mixture density. 

Since the portfolio score in Definition~\ref{def:score} is equal to the expectation of an indicator function with respect to the measure induced by a mixture of log-concave distributions, it can not be computed exactly (e.g. from a closed-form). In the sequel, we discuss how we can estimate the value of the new score by approximating the value of the corresponding multivariate integral.

\subsection{Computation of the score}\label{subsec:computing_the_score}

This section provides a Markov Chain Monte Carlo (MCMC) integration method to guarantee fast and robust approximation within arbitrarily small error for the score in Definition~\ref{def:score}. Let the probability density function $\pi(x) =\sum_{i=1}^Mw_i\pi_i(x)$ to be a mixture of log-concave densities (i.e.\ $\pi_i$ are log-concave distributions). Furthermore, let the vector of assets' returns $R\in\RR^n$, the halfspace $H(R^*):=\{ x\in\RR^n\ |\ R^Tx\leq R^* \}$ and the indicator function $g(x) = \left\{
\begin{array}{ll}
      1. & \mbox{ if } x\in H(R^*) ,\\
      0, & \mbox{otherwise.}\\
\end{array} 
\right.$. Then the score in Equation~(\ref{eq:new_score}) can be written,
\begin{equation}\label{eq:integrals}
\begin{split}
s & = \int_{P}g(x)\sum_{i=1}^Mw_i\pi_i(x)dx = \sum_{i=1}^Mw_i\int_{P}g(x)\pi_i(x)dx \\ & = \sum_{i=1}^Mw_i\int_{P\cap H(R^*)}\pi_i(x)dx  = \sum_{i=1}^Mw_i\int_{S}\pi_i(x)dx ,
\end{split}
\end{equation}
where $S:=P\cap H(R^*)$.

It is clear that the computation of the score $s$ is reduced to integrate $M$ log-concave functions over a convex set $S$, i.e.\ to compute each $\int_{S}\pi_i(x)dx,\ i\in[M]$. For each one of these $M$ integrals, we use the algorithm presented in \cite{Lovasz06} to approximate it within an arbitrarily small error after a number of operations that grows polynomially with the dimension (number of assets) $n$. First, we use an alternative representation of the volume of $S$, employing a log-concave function $\pi(x)$,
\begin{equation}\label{eq:mmc_score}
\begin{split}
& \vol(S) = \int_{S}\pi(x)dx\ \frac{\int_{K}\pi^{\beta_1}(x)dx}{\int_{S}\pi(x)dx}\ \frac{\int_{S}\pi^{\beta_2}(x)dx}{\int_{S}\pi(x)^{\beta_1}dx}\ \cdots\ \frac{\int_{S}1dx}{\int_{S}\pi(x)^{\beta_k}dx}\\
& \Rightarrow \int_{S}\pi(x)dx = \vol(S)\ \frac{\int_{S}\pi(x)^{\beta_k}dx}{\int_{S}1dx}\ \cdots\ \frac{\int_{S}\pi(x)dx}{\int_{S}\pi(x)^{\beta_1}dx} ,
\end{split}
\end{equation}
where the sequence $\beta_j,\ j\in[k]$ are factors applied on the variance of $\pi(x)$.

Since $S$ is the intersection of a halfspace with the convex polytope $P$ we use the algorithm in~\cite{CousinsV14} to approximate $\vol(S)$ within error $\epsilon$ after $O^*(n^3)$, where $O^*(\cdot)$ suppresses polylogarithmic factors and dependence on $\epsilon$. In the special case of $P=\Delta^{n-1}$, we can compute the exact value of $\vol(S)$ using Varsi's algorithm~\cite{Varsi73} after $n^2$ operations at most. Consequently, the computation of $\int_{S}\pi(x)dx$ is reduced to compute $k$ ratios of integrals. For each ratio we have,
\begin{equation}
\begin{split}
r_j = \frac{\int_{S}\pi(x)^{\beta_{j-1}}dx}{\int_{S}\pi(x)^{\beta_j}dx} & = 
\frac{1}{\int_S \pi(x)^{\beta_j}dx}\int_S\frac{\pi(x)^{\beta_{j-1}}}{\pi(x)^{\beta_j}(x)}\pi(x)^{\beta_j}(x)dx \\
& = \int_S\frac{\pi(x)^{\beta_{j-1}}}{\pi(x)^{\beta_j}}\frac{\pi(x)^{\beta_j}}{\int_S \pi(x)^{\beta_j}dx}dx .
\end{split}
\end{equation}

Thus, to estimate $r_j$ we just have to sample $N$ points from the distribution proportional to $\pi(x)^{\beta_j}$ and restricted to $S$. Then,
\begin{equation}\label{eq:ri}
    r_j\approx \frac{1}{N}\sum_{i=1}^N\frac{\pi(x_i)^{\beta_{j-1}}}{\pi(x_i)^{\beta_j}}
\end{equation}
as $N$ grows. The key for an efficient approximation of $r_j$ using Monte Carlo integration is to set $\beta_j,\ \beta_{j+1}$ such that the variance of $r_j$ is as small as possible (ideally a constant) for $N$ as small as possible. In particular, In \cite{Lovasz06}, they prove that the sequence of $\beta_1,\dots ,\beta_k$ can be fixed such that the variance of each $r_j,\ j\in[k]$ is bounded by a constant. Moreover, $N=O^*(\sqrt{n})$ points per integral ratio $r_j$ and $k=O^*(\sqrt{n})$ ratios in total suffices to approximate each $\int_{S}\pi_i(x)dx, i\in[M]$ within error $\epsilon$. Thus, $O^*(n)$ points suffices to estimate each $\int_{S}\pi_i(x)dx$. 
%

\begin{lem}
Let the probability density function $\pi(x)$ in the Definition~\ref{def:score} be a mixture of $M$ log-concave densities. The integral ratio in Equation~(\ref{eq:ri}) can be estimated with $O^*(n)$ samples from $\pi(x)^{\beta_j}$ within error $\epsilon$. Thus, the portfolio score in Equation~(\ref{eq:new_score}) can be estimated using $O^*(Mn)$ samples.
\end{lem}

To sample from each target distribution proportional to $\pi(x)^{\beta_j}$ and restricted to $S$, in~\cite{Lovasz06} they use Hit-and-Run random walk \cite{Vempala07}. This implies a total number of $O^*(n^4)$ arithmetic operations per generated point. Thus the total number of arithmetic operations to estimate the score $s$ is $O^*(Mn^5)$. 
In our implementation, to sample from a log-concave distribution supported on $P$, we use the reflective Hamiltonian Monte Carlo in~\cite{afshar2015reflection} which is more efficient in practice than Hit-and-Run. 
For an extended introduction to geometric random walks we suggest \cite{Vempala07}.

\section{Mixed strategies}\label{sec:determine_parameters}

An important question is how one could set the risk and dispersion parameters $q_i,\ \alpha_{ij}$ and the weight $w_{ij}$ of each allocation strategy $F_{\pi_{q_i,\alpha_{ij}}}$ in a certain stock market. The issue is that our knowledge about the stock market and the behavior of the investors in it might be weak or vary from a time period to another. 

In this section, we provide practical methods to set the parameters of a sequence of log-concave distributions. We also present different versions of the score than those given in Definition~\ref{def:score}. We detail the computational methods of this section in Appendix~\ref{appnd:computational_methods}.

\subsection{Set the levels of dispersion}\label{subsec:compute_a_sequence}

Let the concave function $h(x):P\rightarrow \RR$, where $P\subset\RR^n$ the set of portfolios. Also let the log-concave probability density function,

\begin{equation}
    \pi_{\alpha}(x) \propto e^{\alpha h(x)},\ \alpha > 0 ,
\end{equation}
supported also on $P$ and $\tilde{x}\in P$ the mode of $\pi_{\alpha}$. 
Recall that small/large values of $\alpha$ correspond to large/small values of variance of $\pi_{\alpha}$. Thus, first we compute a value $\alpha_L$ such that $F_{\pi_{\alpha_L}}$ is a $e$-dispersed allocation strategy; that is the distribution $\pi_{\alpha_L}$ is e-close to the uniform distribution according to the $L_2$ norm. Second, we compute a value $\alpha_U$ such that the mass of the distribution $\pi_{\alpha_U}$ is almost entirely concentrated in a ball  $B(\tilde{x},\delta)$, that is a ball centered at the mode $\tilde{x}$ and with a small radius $\delta > 0$. Then, we compute a sequence $\alpha_L = \alpha_1 < \dots < \alpha_{k} = a_U$ such that,

\begin{equation}\label{eq:equidistant_sequence}
    \| \pi_{\alpha_{i+1}} / \pi_{\alpha_i} \| = \| \pi_{\alpha_i} / \pi_{\alpha_{i-1}} \|,\ i \in \{ 2,\dots , k-1 \} .
\end{equation}

To compute $\alpha_L$ we start with $\alpha_0 = 1$ and we use the annealing schedule in~\cite{CousinsV14}. In particular we generate the sequence,

\begin{equation}
    \alpha_i = \alpha_0 \bigg( 1 - \frac{1}{n} \bigg)^i,\ i\in\mathbb{N_+} .
\end{equation}
This schedule guarantees that a sample from $\pi_{\alpha_{i}}$ is a warm start to sample from $\pi_{\alpha_{i+1}}$ for several random walks~\cite{Lovasz06,CousinsV14} and moreover, the variance of the distribution which is proportional to $e^{(\alpha_{i+1}-\alpha_i)h(x)}$ is $O(1)$; that is, each jump to the next distribution in the sequence is "small". 
Next, for each $\alpha_i$ we estimate the $L_2$ norm of $\pi_{\alpha_i}$ w.r.t.\ the uniform distribution, by sampling from $\pi_{\alpha_i}$. We stop when the norm is smaller than a given threshold. 

To compute $\alpha_U$ we use the same annealing schedule, but now we generate an increasing sequence,

\begin{equation}
    \alpha_i = \alpha_0 \bigg( 1 + \frac{1}{n} \bigg)^i,\ i\in\mathbb{N_+} .
\end{equation}
We stop when we meet the smallest $i$ such that the $100(1-\epsilon)\%$ of the mass of $\pi_{\alpha_i}$ is inside the ball $B(\tilde{x},\delta)$ with high probability. We probabilistically guarantee this by sampling a sufficiently large number of points from $\pi_{\alpha_i}$ and by splitting the sample to $\nu$ sub-samples. For each sub-sample we compute the ratio of points that lie in $B(\tilde{x},\delta)$; that is we obtain $\nu$ ratios. Then, we perform a t-test using those ratios while the null hypothesis states that the overall ratio is larger than $(1-\epsilon)$. We stop for an $\alpha_i$ that results in rejecting the null hypothesis.

Finally, to compute a sequence of equidistant distributions as in Equation~(\ref{eq:equidistant_sequence}), we estimate $d = \max\limits_{i\in[k-1]}\{ \| \pi_{\alpha_{i+1}} / \pi_{\alpha_i} \|\}$. Then, we start from $\alpha_1 = \alpha_L$. Given $\alpha_i$, to compute the next value of parameter in the sequence, namely $\alpha_{i+1}$, we perform bisection method in the interval $[\alpha_{i}, \alpha_U]$ to compute a value such that the $L_2$ norm of $\pi_{\alpha_{i+1}}$ w.r.t.\ $\pi_{\alpha_i}$ is $d\pm\epsilon$ with a high probability and a small $\epsilon > 0$. We stop when we compute an $\alpha_i > \alpha_U$ and we set $\alpha_k = \alpha_i$. To select $M$ values of $\alpha$ we pick $\alpha_1$ and $\alpha_k$ and then, we equidistantly pick $M-2$ values in between them.

\subsection{Set the levels of risk}\label{subsec:compute_q_sequence}

Our practical method computes a sequence $q_1<\dots <q_M$. The values $q_i$ are equidistant concerning the portfolio volatility that each $q_i$ corresponds to. 
First, we compute the minimum and the maximum value of portfolio volatility. The first one is also called Global Minimum Variance portfolio~\cite{Zhao20}. In particular, we solve the following optimization problems,

\begin{equation}
    \min/\max\ x^T\tilde{\Sigma} x^T,\ x\in P,\ \tilde{\Sigma}\in\RR^{n\times n}\text{ pos. def.}
\end{equation}
where $P$ is the set of portfolios and $\tilde{\Sigma}$ is an estimation of the covariance using the shrinkage estimate in~\cite{LW04}. Let the values of the minimum and the maximum portfolio volatility $v_{\min}$ and $v_{\max}$ respectively. To compute $M$ values of the parameter $q$, we equidistantly select $M$ values of portfolio volatility $v_{\min} < v_1 < \dots <v_M < v_{\max}$. Then, for each $v_i$ we perform a bisection method in a proper interval $[q_{\min},q_{\max}]$ to compute a $q_i$ such that, 
\begin{equation}
    | \min\limits_{x\in P} \phi_{q_i}(x) - v_i | \leq \epsilon ,\ i\in[M] ,
\end{equation}
for a sufficiently small value of $\epsilon > 0$, while $\phi_q(x)$ is the expected quadratic utility function in Equation~(\ref{eq:mean_variance_function}). In particular, for each $q_i$ we search in $[q_{i-1},q_M],\ i \in\{2,\dots ,M-1\}$; for $q_1$ we search in $[0,q_M]$. To compute $q_M$ we search for the smallest non-negative integer $j$ such that $\min\limits_{x\in P} \phi_{2^j}(x) > v_M$. Then, we perform a bisection method in $[0, 2^j]$ to compute $q_M$. 

\subsection{Set the composition of the investors}\label{subsec:composition_of_investors}

The computation of both sequences of $q$ and $\alpha$ allow to specify the sequence of log-concave distributions, 
\begin{equation}
\pi_{ij}= e^{-\alpha_{ij}\phi_{q_i}(x)},\ i\in[M_1],\ j\in[M_2] ,
\end{equation}
where we assume that for each level of risk $q_i$ we have $M_2$ levels of dispersion. 
However, to determine a mixed strategy one has to determine the weights $w_{ij}$ in the corresponding mixture distribution. We recall that each $w_{ij}$ implies the proportion of investors that build their portfolios according to the allocation strategy induced by $\pi_{ij}$. Setting $w_{ij}$ forms the mixed strategy $F_{\pi}$ while the score in Definition~\ref{def:score} becomes, 
\begin{equation}\label{eq:score2}
s =  \sum_{i=1}^{M_1}\sum_{j-1}^{M_2}w_{ij}\int_{S}\pi_{ij}(x)dx,\ S:=P\cap H(R^*) .
\end{equation}

First, we allow setting additional bounds on $w_{ij}$. For example, one would provide an upper/lower bound on the proportion of the investors who chose a specific allocation strategy. 
In particular, let us assume that we estimate the $M = M_1 M_2$ integrals of Equation (\ref{eq:score2}) as described in Sec.~\ref{subsec:computing_the_score}. $M$ is the total number of allocation strategies in a certain stock market. Then, let the $M$ values to form a vector $c\in\RR^M$. Also let the corresponding weights $w_{ij}$ in Equation~(\ref{eq:score2}) to form a vector $w\in\RR^M$ such that the score,
\begin{equation}
    s = \langle c, w\rangle ,
\end{equation}
where $\langle \cdot , \cdot \rangle$ denotes the inner product between two vectors. Given a matrix $A\in\RR^{N\times M}$ and a vector $b\in\RR^N$, let the following feasible region of weights,
\begin{equation}\label{eq:feasible_weights}
Q = \left \{ w\in\RR^M\ \bigg|\ Aw\leq b,\ w_i\geq 0,\ \sum_i^M w_i = 1 \right \}\subset \RR^M
\end{equation}
The matrix $A$ and the vector $b$ used to express $N$ further constraints on the weights (e.g.\ lower, upper bounds or any linear constraint on $w_{ij}$).
Notice that if no further constraints are given on the weights, then the feasible region $Q$ is the canonical simplex $\Delta^{M-1}$.

Now let us define three new versions of the score $s$ in Equation~(\ref{eq:score2}).
 
\begin{mdframed}
Let the weights $w\in Q$, where $Q\subset\RR^{M}$ the feasible region in Equation~(\ref{eq:feasible_weights}).
\begin{enumerate}
    \item \textbf{min score}, $s_{\min} :=\min \langle c, w\rangle,\ \text{subject to }Q$.
    \item \textbf{max score}, $s_{\max} :=\max \langle c, w\rangle,\ \text{subject to }Q$.
    \item \textbf{mean score}, $\bar{s} :=\frac{1}{\vol(Q)}\int_Q \langle c, w\rangle\ dw$.
\end{enumerate}
\end{mdframed}

For the scores $s_{\min}$ and $s_{\max}$ one has to solve a linear program for each one of them. The score $\bar{s}$ requires the computation of an integral which can be computed with MCMC integration employing uniform sampling from $Q$; otherwise, it can be reduced to the computation of the volume of a convex polytope since $\langle c,w \rangle$ is a linear function of $w$ with the domain being the set $Q$.  

Let $w_1\in Q$ such that the min score $s_{\min} = \langle c,w_1\rangle$. The weights denoted by the vector $w_1$ imply the proportions of the investors that select each allocation strategy such that the portfolio score $s$ takes its possibly minimum value. Similarly, the vector of weights $w_2\in Q$ such that the max score $s_{\max} = \langle c,w_2\rangle$, implies the proportions of the investors that select each allocation strategy such that the portfolio score $s$ takes its possibly maximum value. Moreover, it is easy to notice that the mean score $\bar{s} = \langle c,\bar{w}\rangle$, while the vector of weights $\bar{w}$ is the center of mass of $Q$. For example, if $Q=\Delta^{M-1}$ (i.e.\ the case where no further constraints are given on the weights) the vector $\bar{w}$ is the equally weighted vector.

However, one may have additional knowledge on how the investors tend to behave in a certain stock market, i.e.\ which allocation strategies they tend to select. We also allow for these degrees of freedom by providing the notion of \textit{behavioral functions} in our context.

\subsubsection{Behavioral functions}\label{subsubsec:param_funs}

In this section, we assume that we are given a set of functions that represents the knowledge, that one may have, related to which allocation strategies the investors tend to select in a certain stock market and time period.
We assume that we are given $M_1 + 1$ functions $f_q, f_{\alpha,i},\ i\in[M_1]$ with the domain being $[0, 1]$ for all of them. We call these functions behavioral functions and we use them to create a vector of weights $w\in\RR^{M}$, that emphasizes specific strategies, where $M=M_1M_2$ is the total number of allocation strategies that take place in the stock market. More specifically, $f_q$ declares the level of risks that the investors tend to select, while $f_{\alpha, i}$ declares the level of dispersion that the investors' portfolios --who select risk $q_i$-- tend to have around the formal allocation proposal.
\begin{figure}[t] \centering
\includegraphics[width=0.3\linewidth]{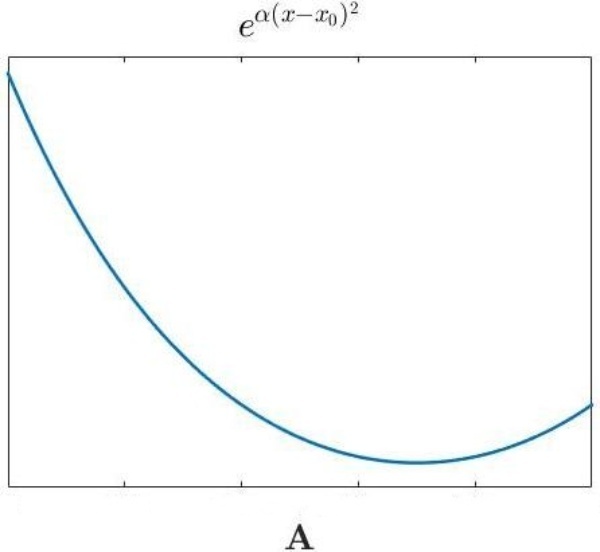}
\hspace{0.5cm}
\includegraphics[width=0.3\linewidth]{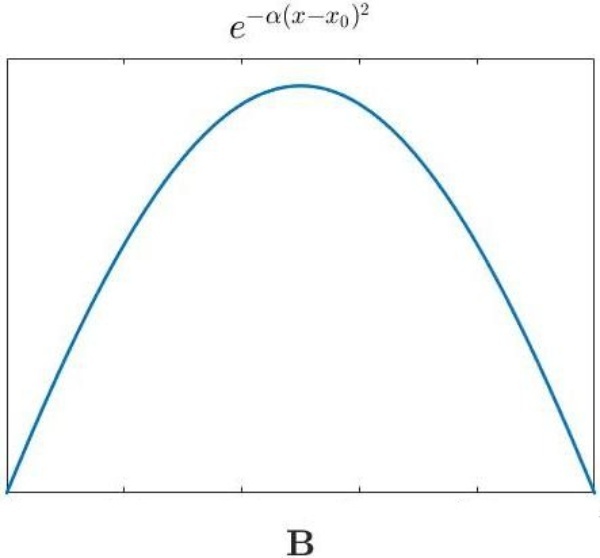}\\
\vspace{0.2cm}
\includegraphics[width=0.3\linewidth]{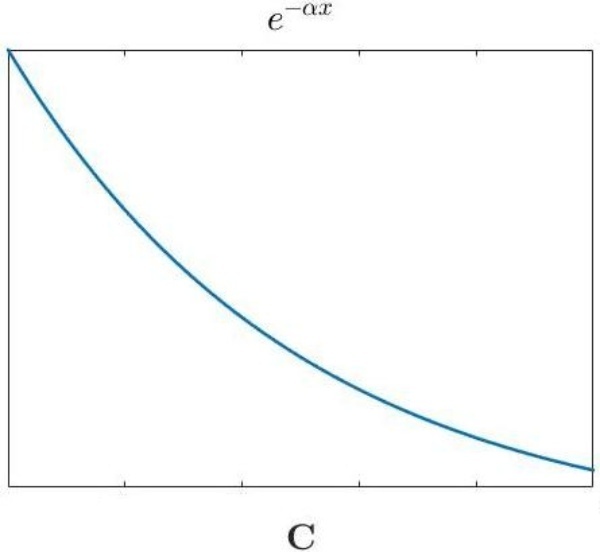}
\hspace{0.5cm}
\includegraphics[width=0.3\linewidth]{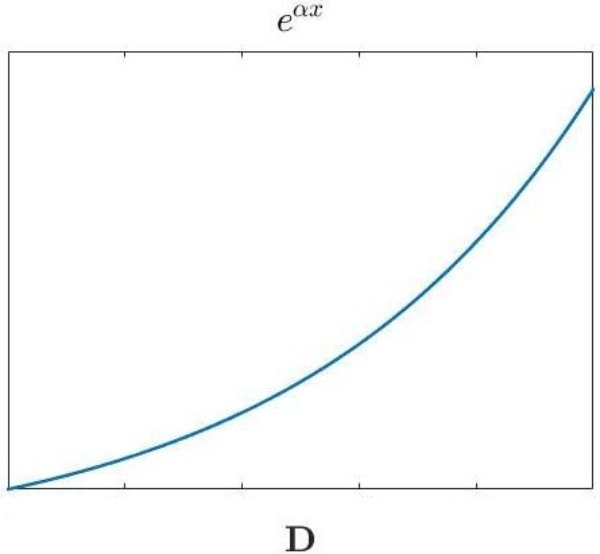}
\caption{Examples of behavioral functions. \label{fig:parameterization_functions}}
\end{figure}

The plots in Fig.~\ref{fig:parameterization_functions} demonstrate $4$ possible choices of such functions. For example, if plot $C$ is $f_q$ then the investors tend to select low-risk investments; the value of $f_q$ is high for small values of $q$ (low risk) and low for high values of $q$ (high risk). In addition, if the plot $D$ is $f_{\alpha, i}$ then, the portfolios of the investors who select risk $q_i$ tend to be highly stuck around the formal allocation proposal that corresponds to $q_i$; the value of $f_{\alpha, i}$ is large for large values of $\alpha$ (low dispersion) and small for small values of $\alpha$ (high dispersion). 

To compute a weight vector $w$ we map the intervals $[a_{i1}, \alpha_{iM2}]$ and $[v_{\min}, v_{\max}]$ we compute in Sec.~\ref{subsec:compute_a_sequence} and Sec.~\ref{subsec:compute_q_sequence} onto $[0,1]$, by using the following transformation:

\begin{equation}
    z(t) = \frac{1}{d-c}(t-c),\ t\in[c,d] .
\end{equation}
Throughout this paper, when we write $z(\cdot )$ we assume that the interval $[c,d]$ is defined properly according to the input.

The following pseudocode describes how we compute such a weight vector when $M_1+1$ behavioral functions are given. 

\begin{mdframed}
\textbf{Compute the weight vector $w$}\\
\textbf{Input}: risk and dispersion parameters $q_i$ and $\alpha_{ij}$, $i\in[M_1],\ j\in[M_2]$\\
\hspace*{1.2cm}computed as in Sec.~\ref{subsec:compute_a_sequence} and~\ref{subsec:compute_q_sequence} and $M_1+1$ behavioral\\ \hspace*{1.1cm} functions$f_q,\ f_{\alpha,i}$.
\begin{enumerate}
    \item Let the portfolio volatilities $v_i = \min\limits_{x\in P}\phi_{q_i}(x),\ i\in[M_1]$.
    \item For each pair of $(i,j)$ set $r_{(i-1)M_1 + j} \leftarrow f_q(z(v_i))f_{\alpha, i}(z(\alpha_{ij}))$.
    \item Normalize the vector $r_j\leftarrow r_j/\sum_{i=1}^Mr_i,\ j\in[M]$ and $M=M_1M_2$.
    \item Set the weight vector $w\leftarrow r$.
\end{enumerate}
\end{mdframed}

Given the behavioral functions, one could use the vector of weights ---determined as in the above pseudo-code--- and then, the portfolio score is $s = \langle c,w\rangle$,   
while $c\in\RR^M$ is the vector that contains the values of the integrals in Equation~(\ref{eq:score2}).

\subsubsection{Parametric score}\label{subsubsec:parametric_score}

In this section, we assume a weaker knowledge of how the investors tend to behave than that in Sec.~\ref{subsubsec:param_funs}. Thus, we do not explicitly determine the vector of weights $w\in\RR^M$, where $M$ is the number of allocation strategies in a certain stock market. In particular, let the coordinates of the vector $r\in\RR^M$ as in Sec.~\ref{subsubsec:param_funs}, 
\begin{equation}\label{eq:bias_vector}
r_{(i-1)M_1 + j} \leftarrow f_q(z(v_i))f_{\alpha, i}(z(\alpha_{ij})),\ i\in[M_1],\ j\in[M_2]
\end{equation}
where $f_q,\ f_{\alpha, i}$ the $M_1+1$ behavioral functions. Then, we use the vector $r$ to denote a {\em bias} on the investors' behavior. First, we again allow further bounds and linear constraints on the weights. That is, we let the feasible region of the weights to be the set $Q$ in Equation~(\ref{eq:feasible_weights}). To denote the bias on the investors's behavior, we employ the exponential distribution, 
\begin{equation}
p_T(w)\propto e^{\langle r, w\rangle /T} ,\ T>0 ,
\end{equation}
with the support of $p_T(w)$ being the set $Q\subset\RR^M$. 

The distribution $p_{T}(w)\propto e^{\langle r, w\rangle /T}$ is usually called Boltzmann distribution and the vector $r$ {\em bias vector}. In general, the Boltzmann distribution gives the probability that a system will be in a certain state as a function of that state's energy and the temperature $T$ of the system. The bias vector $r$ determines how the mass tends to distribute in $Q$ and the temperature parameter $T$ how strong the bias denoted by $r$ is. 

For example, when $Q$ is the canonical simplex $\Delta^{M-1}$, the mass of $p_T$ tends to concentrate around the vertices which correspond to the coordinates of $r$ with larger values than the other coordinates. Moreover, as the temperature $T\rightarrow 0$ this tendency becomes stronger until almost all the mass concentrates around the vertex which corresponds to the coordinate of the largest value of $r$. As $T\rightarrow\infty$, $p_{T}$ converges to the uniform distribution and the bias denoted by $r$ disappears. 

We intend to use the temperature $T$ to parameterize how strong the tendency on the investors' behavior, that the bias vector $r$ implies, is. Then the parametric score is given as,
\begin{equation}\label{eq:parametric_score}
\begin{split}
     & s(T) := \int_{S} \langle c, w\rangle\ p_{T}(w) dw,\ \text{where }p_{T}(w) \propto  e^{\langle r, w\rangle /T},\ T>0\\ &\text{ and each coordinate }\quad r_{(i-1)M_1 + j}= f_{q}(z(v_i))f_{\alpha, i}(z(\alpha_{ij})),\\ & \text{ and } v_i = \min\limits_{x\in P}\phi_{q_i}(x),\ i\in[M_1],\ j\in[M_2]
\end{split}
\end{equation}


Let the center of mass $\bar{w_T}$ in $Q$ when the mass is distributed according to $p_T(w)$. Notice that $\bar{w}_T$ can be seen as a parametric curve on $T$. Furthermore, it is easy to notice that, for fixed $T$, the parametric score $s(T) = \langle c,\bar{w}_T\rangle$. Thus, the score $s(T)$ is evaluated on that parametric curve. Following these observations, we are ready to state the following Lemma.

\begin{lem}\label{lem:scores}
Let a stock market with $M$ allocation strategies. 
Assume that we are given the parameters $q_i,\ \alpha_{ij}$ of Sec.~\ref{subsec:compute_a_sequence} and~\ref{subsec:compute_q_sequence} and any behavioral functions $f_q,\ f_{\alpha_i},\ i\in[M_1],\ j\in[M_2]$ and $M=M_1M_2$ the number of allocation strategies that take place in the stock market. Let the feasible set $Q\subset\RR^{M}$ of the weights as in Equation~(\ref{eq:feasible_weights}), the min score $s_{\min}$, the max score $s_{\max}$ and the mean score $\bar{s}$ in Sec.~\ref{subsec:composition_of_investors} and the parametric score in Equation~(\ref{eq:parametric_score}). Then, the followings hold,
\begin{equation}\label{eq:lem_scores}
\begin{split}
    & s_{\min}\leq s(T) \leq s_{\max},\ \forall T>0,\\
    & \bar{s} = \lim_{T\rightarrow \infty}s(T)
\end{split}
\end{equation}
\end{lem}

Notice that the Equation~(\ref{eq:lem_scores}) holds for any set of behavioral functions as the scores $s_{\min},\ s_{\max}$ always bound the parametric score. 
Furthermore, when $T\rightarrow \infty$ the distribution $p_T(w)$ converges to the uniform distribution over the feasible region of the weights $Q$ and thus the parametric score is equal to the mean score $\bar{s}$. 

To obtain the parametric score we compute a sequence of temperatures $T_i$ that correspond to a sequence of exponential distributions $p_{T_i}$. Similarly to Sec.~\ref{subsec:compute_a_sequence}, we compute two temperatures $T_{\max}$ and $T_{\min}$. The $L_2$ norm of $p_{T_{\max}}$ w.r.t.\ the uniform distribution over $Q$ is smaller than a given threshold and the $100(1-\epsilon)\%$ of the mass of $p_{T_{\min}}$ is inside a ball of a small radius $\delta > 0$, centered at the mode of $p_{T_{\min}}$. Then, we use the sequence, $T_i = T_{\max}(1 - \frac{1}{M})^i$, $i\in\mathbb{N}_+$, $T_i\geq T_{\min}$ and the method in Sec.~\ref{subsec:compute_a_sequence} to compute an equidistant --with respect to $L_2$ norm-- sequence of exponential distributions.

\section{Simulations on allocation strategies}\label{sec:simulations}

In this section, we take the set of portfolios $P\subset\RR^n$ to be the canonical simplex, $\Delta^{n-1}$, which means that we consider long-only portfolios. 
We illustrate the usefulness of the new score in analyzing the performance of a portfolio allocation given the asset returns. We also compare it to several well-known portfolio scores. Moreover, we consider the score of a given portfolio as a random variable in a stock market, where the asset returns follow a multivariate distribution. We also illustrate how our modeling of allocation strategies could be used to study the state of a stock market, computing the copula between portfolios' return and volatility while the portfolios have been build according to a mixed strategy.

\subsection{Portfolio score}\label{subsec:illustrate_score}

In our simulation, we consider the daily returns of the $12$ cryptocurrencies with the longest history, reported in Table~\ref{Tab:cryptos}. To illustrate our new score, we consider a pseudo-real time example, where we take $100$ consecutive asset returns from 22/10/2016 until 29/01/2017. We compute the optimal Mean-Variance (MV) portfolio using the shrinkage estimate of the covariance matrix of~\cite{LW04}, while we fix its volatility equal to the average in-sample volatility of the long-only portfolios. We also compute the Equally-weighted Risk Contributions (ERC) portfolio by~\cite{MRT2010} which also uses the shrinkage estimator and the Bitcoin (BTC) portfolio. For the sake of completeness we report the estimated covariance matrix and the average assets' return for the period of $100$ days, that we used to compute the MV portfolio, in Appendix~\ref{appnd:crypto_data}. We also report the three portfolios in Table~\ref{tab:portfolios}.

\begin{table}[h!]
\centering
\begin{tabular}{c|ccccccc}
Portfolios & BTC    &   LTC    &  ETH    &  XRP    &    XMR  &  USDT \\
\hdashline
MV & 0\%  & 58.5\%  &  0\% &    0\% &    0\% &    0\%  \\
 ERC & 8.94\%  &  5.81\% & 6.81\%  &  16.18\%  &  7.02\% & 7.70\% \\
 BTC & 100\%  &  0\% & 0\%  &  0\%  &  0\% & 0\% \\
\hline
Portfolios &  DASH  &   XLM    &  DOGE     &  DGB  &  XEM  &  SC  \\
\hdashline
MV  & 0\% &   9.98\% & 0\%  &  1.15\% &   30.39\% & 0\% \\
ERC & 6.38\% & 5.80\% &   8.82\% & 3.88\% &   5.89\% & 16.77\% \\
 BTC & 0\%  &  0\% & 0\%  &  0\%  &  0\% & 0\% \\
\end{tabular}
\caption{The Mean-variance (MV) optimal portfolio, the Equally-weighted Risk Contributions (ERC) portfolio and the Bitcoin portfolio (BTC). 
\label{tab:portfolios}}
\end{table}

To evaluate those portfolios we take the average of the $10$ vectors of assets' returns after the $100$ daily asset returns. We report this vector of assets' returns in~\ref{appnd:crypto_data}. Together with our score, we report Jensen's alpha, Sharpe ratio, Sortino ratio, and the cross-sectional score in~\cite{Guegan11}. The latter is equal to the proportion of all possible allocations that our portfolio outperforms. To compute the Jensen's alpha and the Sortino ratio, we consider the return of the Global Minimum Variance portfolio as the risk-free rate and for the market portfolio, we set the equally weighted portfolio. To compute the Sharpe ratio we also set the equally weighted portfolio as the benchmark portfolio. 

\begin{table}[t]
\centering
\begin{tabular}{c|cccc}
Portfolios &     Jen.\ alpha   &  Sharpe r. & Sortino r. & UnBiasedSc  \\
\hline
MV & -0.0026 &  0.070 &  1.38 &   70.2\%  \\
 ERC &   -0.0014 & -0.16 &  0.86 &  34.8\%   \\
  BTC &-0.014 & -0.41 &  0.015 &   1.8\%  \\
\end{tabular}
\caption{Four well known scores of the Mean-variance (MV) optimal portfolio, the Equally-weighted Risk Contributions (ERC) portfolio and the Bitcoin portfolio (BTC). From left to right: Jensen's alpha, Sharpe ratio, Sortino ratio, the cross sectional score in~\cite{Guegan11}.
\label{tab:values_scores}}
\end{table}

Regarding our new score, we study two scenarios that differ based on the strategies that take place in the stock market. First, we take $3$ levels of risk with $4$ levels of dispersion for each risk, that is $M=12$ strategies in total. Second, we take $6$ levels of risk with $10$ level of dispersion for each risk. In addition, we select $10$ levels of dispersion around the Bitcoin portfolio. More precisely, consider the family of distributions, 

\begin{equation}
    \pi(x) \propto e^{-\alpha (x-\bar{x})^T\Sigma(x-\bar{x})},\ x\in\RR^n ,
\end{equation}
where $\bar{x}\in\RR^n$ is the Bitcoin portfolio. Then, we compute the sequence of dispersion as in Sec.~\ref{subsec:compute_a_sequence}. That is $M=70$ strategies in total. In both cases, we do not impose any additional constraint for the proportion of the investors that select a specific strategy. This means that the set of weights $Q\subset\RR^M$, which determines the composition of the investors in the stock market, is the canonical simplex $\Delta^{M-1}$. Considering the behavioral functions, for the risk we consider three cases: (i) plot B with $x_0=1/2$, (ii) plot C and (iii) plot D in Fig.~\ref{fig:parameterization_functions}. The function in case (i) favors strategies with medium level of risk, the function in case (ii) favors strategies with low level of risk, and (iii) favors strategies with high level of risk. Throughout this section for the dispersion of each risk, we consider only the case of plot D in Fig.~\ref{fig:parameterization_functions}, which favors strategies of low dispersion around the formal allocation proposal. For all the behavioral functions, we set the ratio between its maximum over its minimum value equal to $10$.

\begin{table}[h!]
\centering
\begin{tabular}{|c|cccc|cccc|}
\hline
 & \multicolumn{4}{c|}{12 Allocation Strategies} & \multicolumn{4}{c|}{70 Allocation Strategies} \\
Portfolios & $\bar{s}$    & $s$ (HR) & $s$ (MR) & $s$ (LR)   &  $\bar{s}$    & $s$ (HR) & $s$ (MR) & $s$ (LR)  \\
\hline
MV & 67.6\%  &  57.4\% & 72.9\% & 82.1\% & 71.2\% & 53.7\% & 71.6\% & 81.0\%  \\
 ERC &  22.1\% & 13.7\%  & 19.8\% &  29.9\% & 33.7\% & 22.7\% & 32.2\% & 41.8\%  \\
  BTC & 0.20\%  &  0.07\% & 0.06\% & 0.08\% & 0.42\% & 0.14\% & 0.21\% & 0.25\% \\
\hline
\end{tabular}
\caption{The scores of the Mean-variance (MV) optimal portfolio, the Equally-weighted Risk Contributions (ERC) portfolio, and the Bitcoin portfolio (BTC) when $12$ and $70$ strategies take place in the stock market. $\bar{s}$ stands for the mean score; $s$ (HR) stands for the score when the behavioral function of the risk is given by plot D in Fig.~\ref{fig:parameterization_functions} (favors high risk strategies); $s$ (MR) stands for the score when the behavioral function of the risk is given by plot B in Fig.~\ref{fig:parameterization_functions} (favors medium risk strategies); $s$ (LR) stands for the score when the behavioral function of the risk is given by plot C in Fig.~\ref{fig:parameterization_functions} (favors low risk strategies). For all cases, the behavioral function of the dispersion is given by plot D in Fig.~\ref{fig:parameterization_functions}, i.e.\ it favors strategies of low dispersion around the formal allocation proposal.
\label{tab:new_score}}
\end{table}

In Table~\ref{tab:values_scores} we report the values of the existing portfolio scores. All scores agree that the performance of the MV portfolio is better than both ERC's and BTC's. They all also agree that ERC's performance is better than BTC's. Moreover, MV is the only portfolio that outperforms the equally weighted portfolio. The cross-sectional score in~\cite{Guegan11} informs us that MV outperforms the $70.2\%$ of all possible portfolios, ERC the $34.8\%$ and BTC outperforms only the $1.8\%$ of all possible portfolios. 
In Table~\ref{tab:new_score} we report the new score when $M=12$ or $M=70$ allocation strategies take place in the stock market. We report the mean score $\bar{s}$, which is the score when the investors are equally divided among the allocation strategies and the scores for the three different choices of the risk's behavioral function, using the weight vector $w$ of Sec.~\ref{subsubsec:param_funs}. 

\begin{table}[t]
\centering
\scriptsize
\begin{tabular}{|ll|rrr|rrr|rrr|}
 \hline
 & & \multicolumn{9}{c|}{Risk's behavioral function} \\
 \hline
 \multicolumn{2}{|c|}{\hspace{-0.6cm}Strategy risk} & \multicolumn{3}{c|}{High risk (Plot D)} & \multicolumn{3}{c|}{Medium risk (Plot B)} & \multicolumn{3}{c|}{Low risk (Plot C)}   \\
  \multicolumn{2}{|c|}{\hspace{-0.6cm}(proposal vol.)} &    $w_1$   &  $w_2$  & $w_3$  & $w_1$   &  $w_2$  & $w_3$     & $w_1$   &  $w_2$  & $w_3$   \\

\hline    
\quad\quad \multirow{ 4}{*}{1.1\%} & $\alpha_{\max}$  & 8.41\%  &  6.14\% & 1.56\% &   10.17\% & 11.47\%  &  3.07\% & 12.27\% &   45.94\% & 85.47\% \\

  & $\dots$ & 7.66\% & 4.66\% &   1.14\% & 7.93\% &  4.36\% & 1.11\% &  8.60\% & 6.08\% & 1.64\% \\
  
 & $\dots$ &  7.45\% & 4.09\% &   1.05\% & 7.33\% &  3.48\% & 0.84\% &   7.80\% &    4.35\% &  1.16\%  \\
 
 & $\alpha_{\min}$  & 7.46\% &  4.11\% &  1.02\% &  7.14\%  & 3.31\% &  0.81\% & 7.60\% &    3.94\% &  1.08\% \\
 
 \hdashline
\quad\quad \multirow{ 4}{*}{1.9\%} &  $\alpha_{\max}$ &   9.78\%   &  10.33\% &   2.65\% &  12.29\% &  42.03\% &  85.31\%  &  9.86\% &   9.47\%  &  2.68\% \\

  & $\dots$ &  8.48\% &  5.57\%  &  1.43\%  &  8.85\%  &  6.20\%  &  1.46\%  &  8.54\%  &   5.31\%  &  1.42\% \\
  
   & $\dots$ & 7.87\%  &  4.42\%  &  1.15\% & 7.62\%  &  3.93\%  &  0.95\% &   7.66\%  &     4.11\%  &  1.13\%  \\
   
   & $\alpha_{\min}$ & 7.54\%  &  4.18\% &   1.03\%  &  7.09\%  &  3.41\%  &  0.85\%  &   7.40\%  &  3.73\%  &  0.99\% \\
   
  \hdashline
\quad\quad  \multirow{ 4}{*}{2.7\%} & $\alpha_{\max}$  &  12.34\% & 43.33\% &   85.63\% & 10.26\%  &  11.70\% & 3.11\%  &  8.47\%  &  5.75\%  &  1.51\%  \\

 & $\dots$ &  7.70\%  &  4.45\%  &  1.14\%  &  7.18\%  &  3.46\%  &  0.85\%  &  7.28\% & 3.84\%   & 0.98\% \\
 
  & $\dots$ &  7.69\%  &  4.40\%  &  1.12\%  &  7.07\%  &  3.34\%  &  0.84\%  &  7.29\%  &  3.77\%  &  0.98\%  \\
  
  & $\alpha_{\min}$ & 7.64\%  &  4.33\%  &  1.09\%  &  7.08\%  &  3.30\%  &  0.80\%  &  7.23\%  &  3.72\%  &  0.96\%  \\
  
\hline\hline
\multicolumn{2}{|c|}{MV score} &  65.4\% &   43.9\%  &  11.1\%   & 68.8\%  &  78.0\% &    94.4\%    &  70.5\%  &  86.2\%  &  96.6\%    \\

\multicolumn{2}{|c|}{ERC score} & 20.9\%  & 12.8\% & 3.2\%  &  21.5\% &  15.1\% & 3.8\% & 23.8\%  & 40.6\%  &  61.7\% \\

\multicolumn{2}{|c|}{BTC score} & 0.18\%  &  0.13\% & 0.02\%  & 0.19\%  &  0.01\% & 0.02\% & 0.18\%  & 0.11\% & 0.02\%  \\
\hline
\end{tabular}
\caption{The scores of the Mean-variance (MV) optimal portfolio, the Equally-weighted Risk Contributions (ERC) portfolio, and the Bitcoin portfolio (BTC) for various weight vectors $w_i$, when 12 strategies take place in the market. Each subset of rows corresponds to a set of allocation strategies with a certain risk level and each row to an allocation strategy with a certain dispersion level. Each weight in $w_i$ gives the percentage of investors who select the corresponding allocation strategy. Each column block corresponds to a different risk behavioral function and a bias vector $r$ according to Eq.~\ref{eq:bias_vector}, each plot refers to Fig.~\ref{fig:parameterization_functions}; each weight vector in a column block corresponds to the center of mass of the distribution $p_T(w)\propto e^{\langle r, w\rangle /T}$ supported on $Q$ for different values of temperature $T$.
For each weight vector, we report the score of the portfolios.
\label{tab:scores_of_portfolios_12_strategies}}
\end{table}

For all portfolios, the score increases while the investors tend to select allocation strategies with s lower level of risk. The performance of the MV portfolio is very similar in both cases of $M=12$ and $M=70$. ERC portfolio performs better when $M=70$ as each score is about $10\%$ larger than the case of $M=12$. BTC also performs better when $M=70$. However, the score is quite small in both cases. Comparing to the unbiased case of the score in~\cite{Guegan11}, it is clear that the value of our score is affected by the investors' composition and can be higher or smaller than the score in~\cite{Guegan11}.

\begin{figure}[t] \centering
\includegraphics[width=0.48\linewidth]{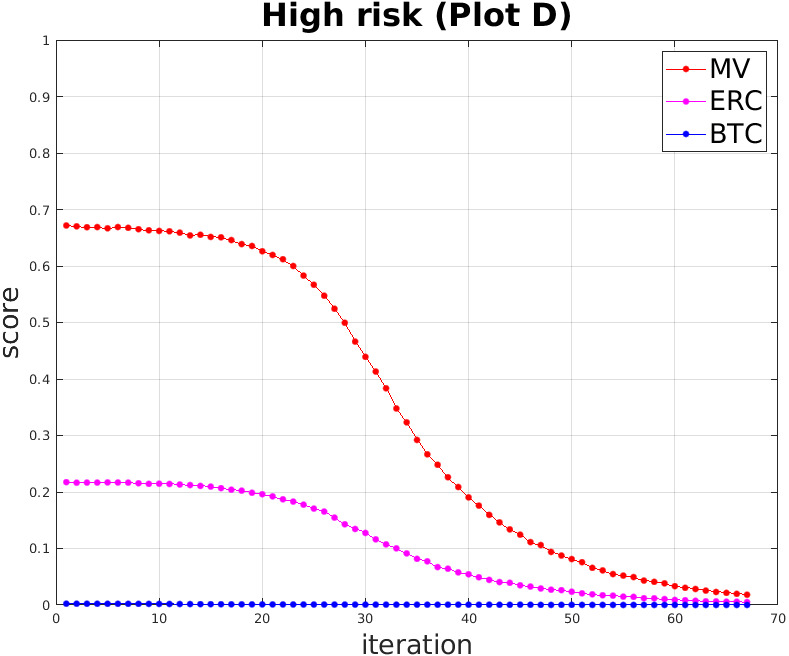}
\includegraphics[width=0.48\linewidth]{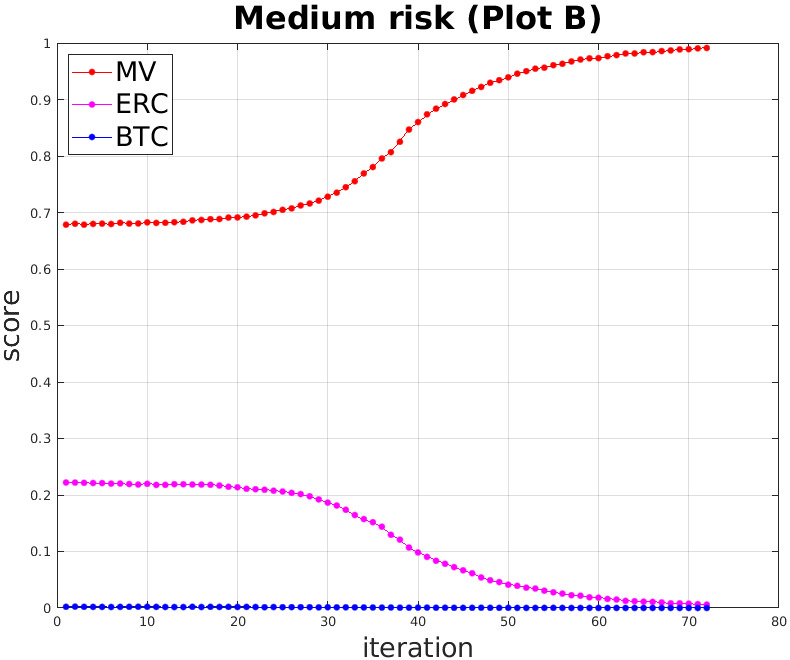}
\includegraphics[width=0.47\linewidth]{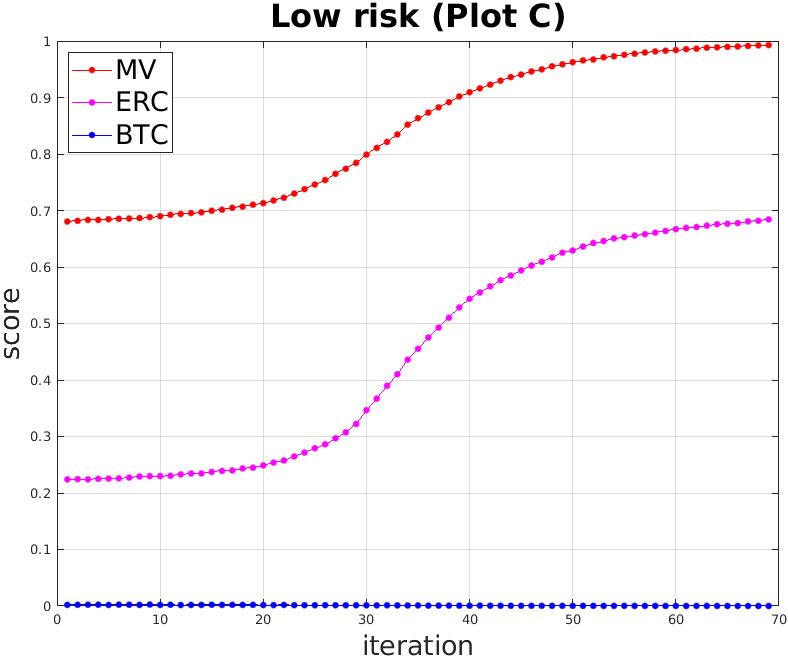}
\caption{The parametric scores of the Mean-variance (MV) optimal portfolio, the Equally-weighted Risk Contributions (ERC) portfolio, and the Bitcoin portfolio (BTC) for different risk's behavioral functions (high, medium, low risk) and $M=12$ strategies in the stock market. We report the score for the $i$-th temperature in the sequence we compute according to Sec.~\ref{subsubsec:parametric_score}. With different colors, we mark different choices for the risk's behavioral function. Plot B, C, D refer to Fig.~\ref{fig:parameterization_functions}. \label{fig:parameteric_scores_12_strategies}}
\end{figure}

In Table~\ref{tab:scores_of_portfolios_12_strategies}, we further illustrate how the performance of each portfolio is related to the investors' composition in the stock market. We consider the case of $M=12$ and for each risk's behavioral function, we compute the bias vector $r$ of Sec.~\ref{subsubsec:parametric_score}. For each bias vector, we report three different weight vectors and the corresponding scores. Recall that a weight vector represents the investor's composition in a stock market. In particular, we set three different temperatures $T$ in the distribution $p_T(w)\propto e^{\langle r, w\rangle /T}$. Then, for each temperature, we estimate the center of mass of $p_{T}$. In each column block and from left to right, we decrease the temperature, and thus, we strengthen the tendency implied by the bias vector $r$. For all portfolios, as the percentage of investors who select strategies with a high level of risk increases, their score drastically decreases. When the percentage of the investors who select strategies with a medium level of risk increases, the performance of the MV portfolio improves while both ERC's and BTC's decrease. When the percentage of investors who select strategies with a low level of risk increases both the MV's and ERC's scores increase while BTC's score decreases. Moreover, BTC's score is always smaller than 0.2\% which implies a quite poor performance.

\begin{figure}[t] \centering
\includegraphics[width=0.48\linewidth]{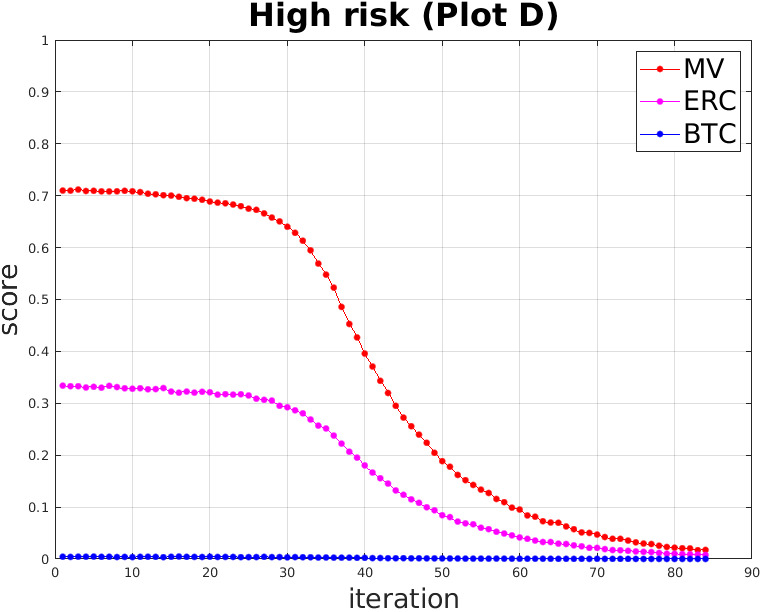}
\includegraphics[width=0.48\linewidth]{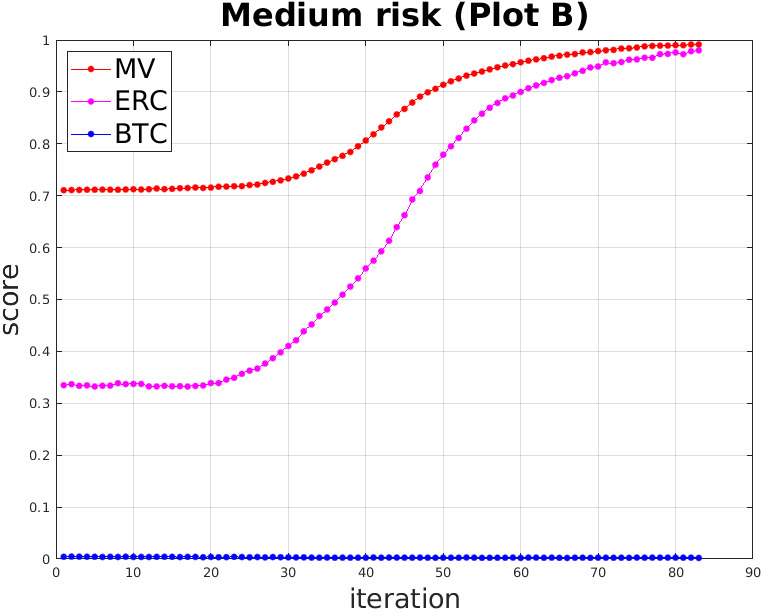}
\includegraphics[width=0.47\linewidth]{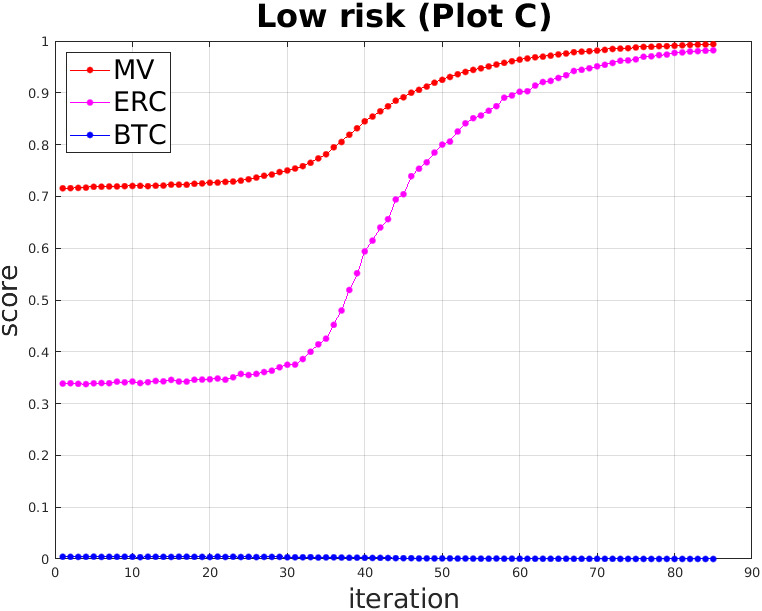}
\caption{The parametric scores of the Mean-variance (MV) optimal portfolio, the Equally-weighted Risk Contributions (ERC) portfolio, and the Bitcoin portfolio (BTC) for different risk's behavioral functions (high, medium, low risk) and $M=70$ strategies in the stock market. We report the score for the $i$-th temperature in the sequence we compute according to Sec.~\ref{subsubsec:parametric_score}. With different colors, we mark different choices for the risk's behavioral function. Plot B, C, D refer to Fig.~\ref{fig:parameterization_functions}. \label{fig:parameteric_scores_70_strategies}}
\end{figure}

The plots in Fig.~\ref{fig:parameteric_scores_12_strategies} and Fig.~\ref{fig:parameteric_scores_70_strategies} illustrate a comparison between the three portfolios using their parametric scores. For both sets of allocation strategies, the score of MV is always higher than the scores of ERC and BTC. When the percentage of investors who select strategies with a high level of risk increases the three parametric scores converge as they all go to zero. When the investors tend to select strategies with a medium or a low level of risk we have a major change in the performance of ERC when $M=70$. For example, for medium risk and $M=12$ the parametric score of ERC converges to 0, while for $M=70$ it converges to 1. This is an example of how the (parametric) score can change as the number of allocation strategies in the stock market also changes.  

\subsection{The score as a random variable}

Once a portfolio is chosen and assuming a distribution for the asset returns, one can estimate the distribution of the scores of this portfolio. This  distribution  allows  us to  understand  the  risk  for  this  portfolio  to  perform  worse,  or better, than a mixed strategy. This estimation is obtained as follows. First, we draw randomly $10^4$ vectors of asset returns. Then we compute the corresponding scores using our implementation. Finally, we estimate the distributions of the score by a normal kernel function bounded in $[0,1]$. Moreover, considering the parametric score we compute a sequence of distributions of scores. That is one distribution for each temperature in Equation~(\ref{eq:parametric_score}).

We show how the mixed strategy in the stock market affects the distribution of the score of a given portfolio. We work under the assumption that the asset returns follow a multivariate Gaussian distribution $\mathcal{N}(\mu,\Sigma)$. We use the same covariance matrix and mean vector as in Sec.~\ref{subsec:illustrate_score}. 

\begin{figure}[t] \centering
\includegraphics[width=0.48\linewidth]{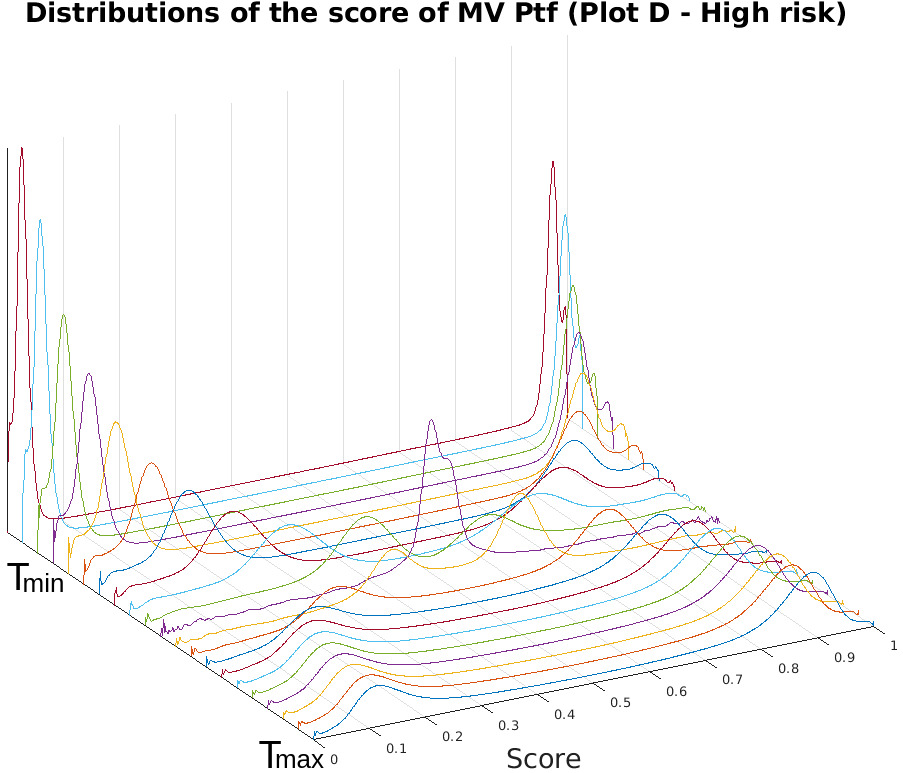}
\includegraphics[width=0.48\linewidth]{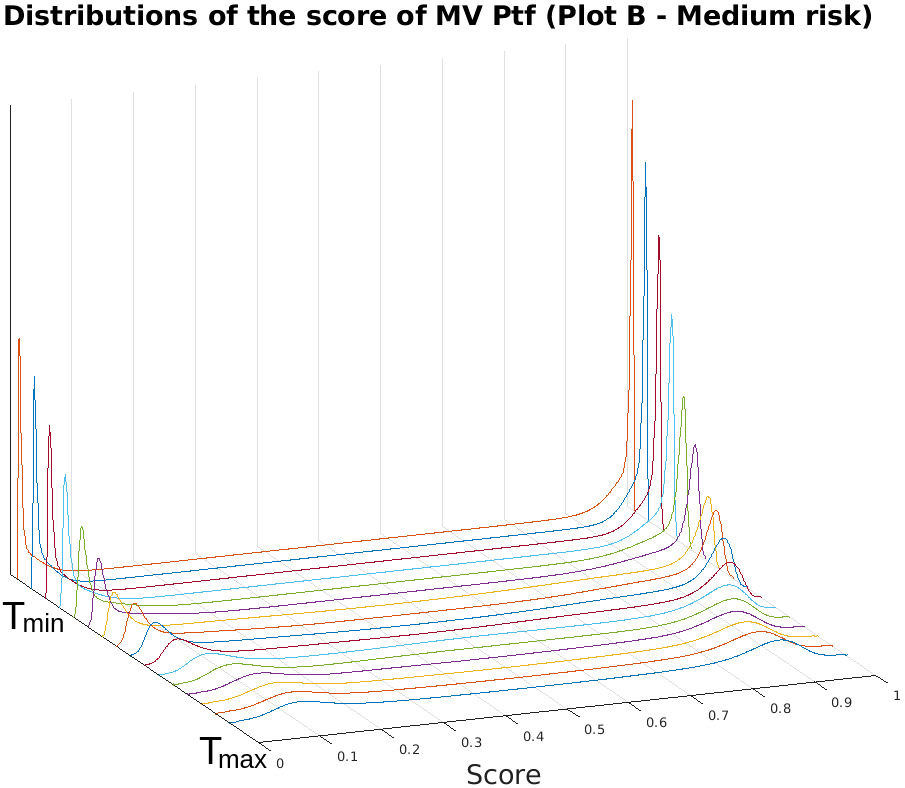}\\
\vspace{0.2cm}
\includegraphics[width=0.47\linewidth]{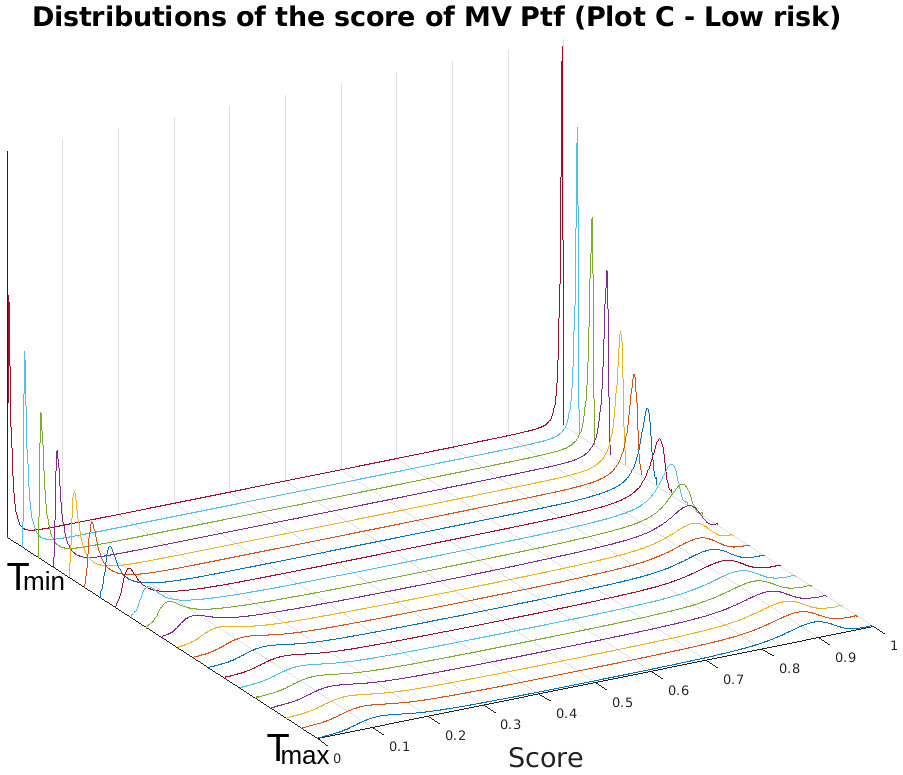}
\caption{The distributions of scores of the Mean-variance (MV) optimal portfolio for different risk's behavioral functions (high, medium, low risk) and a sequence of temperatures $T_i$ such that the distributions in the sequence $p_{T_i}{w}\propto e^{\langle r, w\rangle /T_i}$ are equidistant w.r.t.\ $L_2$ norm. Plot B, C, D refer to Fig.~\ref{fig:parameterization_functions}. \label{fig:parametric_score_distribution}}
\end{figure}

In our example, we focus on the MV portfolio in Table~\ref{tab:portfolios} considering the case of the $M=70$ allocation strategies of Sec.~\ref{subsec:illustrate_score}. For each risk's behavioral function we use the same temperatures as in Fig.~\ref{fig:parameteric_scores_70_strategies}, where we compute the parametric scores. Thus, the plots in Fig.~\ref{fig:parametric_score_distribution} illustrate how the distribution of score of the MV portfolio changes as the tendency of the investors' behavior --induced by the behavioral functions-- increases. 

When the investors are equally divided among allocation strategies, the distribution of score is bimodal with the modes being around $0.2$ and $0.8$. The latter implies that --for this investors' composition-- the score of MV has a high probability to be around 0.8 or 0.2. As the percentage of investors who select strategies with a medium level of risk increases the modes are shifting to the extreme values, e.g. around 0 and 1. Moreover, the total mass over the modes increases. Also, the mass over the mode of the largest value gets larger than the mass over the mode of the smallest value. Thus, the probability of MV to achieve a good score increases as the temperature $T$ decreases. On the other side, the MV has a high probability to be either among the best or worst performers, which implies that it is also quite risky w.r.t.\ to that score. The same occurs as the percentage of investors who select strategies with a low level of risk increases.

However, as the percentage of investors who select strategies with a high level of risk increases, the modes are shifting towards the opposite directions than in the previous cases. Moreover, for the median temperature, the distribution of the score becomes unimodal and it is centered at 0.5. As the temperature further increases the modes are shifting towards the extreme scores, e.g.\ 0 and 1. In this case, the mass over the mode of the largest value gets smaller than the mass over the mode of the smallest value. Thus, as the percentage of investors who select high-risk strategies increases the probability of MV achieving a bad score increases. However, when the investors' composition is implied by the median temperature, MV has a very small probability to be among the best or worst performers, as the mass is concentrated around the score of 0.5. The latter make the MV portfolio a safe (stable) choice w.r.t.\ that score.

\subsection{Alternative Copulae}\label{subsec:alternative_copulas}

Notice that the analysis in Sec.~\ref{sec:crises_detection} is agnostic on allocation strategies by working directly with the set of portfolios. Thus, it uses uniform sampling from the set of portfolios to estimate the copula between portfolios' return and volatility. In this section we compute copulae when the portfolios have been build according to a mixed strategy. 
To be more precise, we consider the case of $M=70$ allocation strategies as in Sec.~\ref{subsec:illustrate_score}, which are computed using the covariance matrix and the mean asset returns estimated on the period of $100$ days of Sec.~\ref{subsec:illustrate_score}. Then, we consider the next $60$ days after the set of $100$ days, and --in that time period-- we compute one copula per mixed allocation strategy. To compute each copula we use sampling from the corresponding mixture distributions --for more details see Appendix~\ref{appnd:computational_methods}.

\begin{figure}[t] \centering
\includegraphics[width=0.35\linewidth]{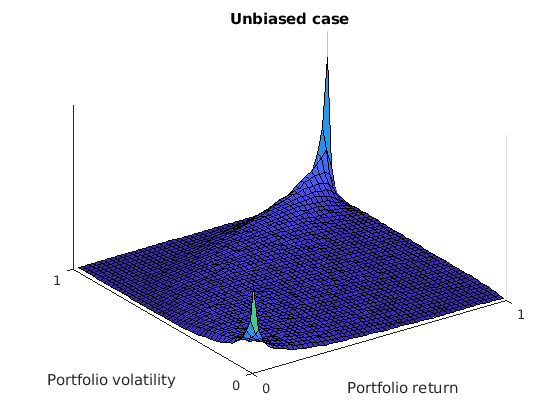}
\includegraphics[width=0.35\linewidth]{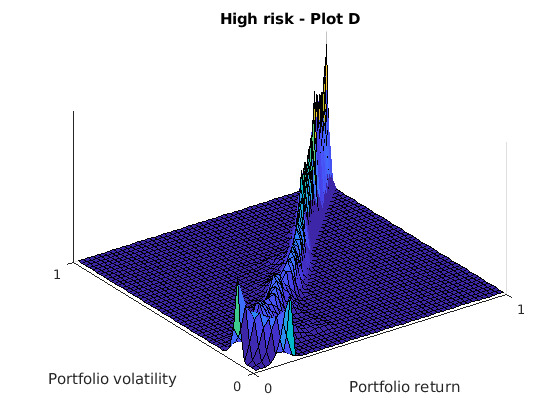}\\
\vspace{0.4cm}
\includegraphics[width=0.35\linewidth]{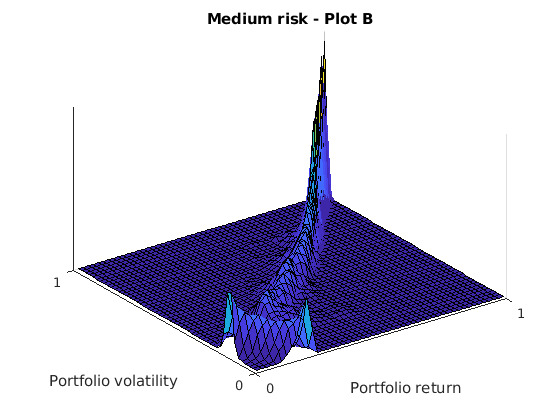}
\includegraphics[width=0.35\linewidth]{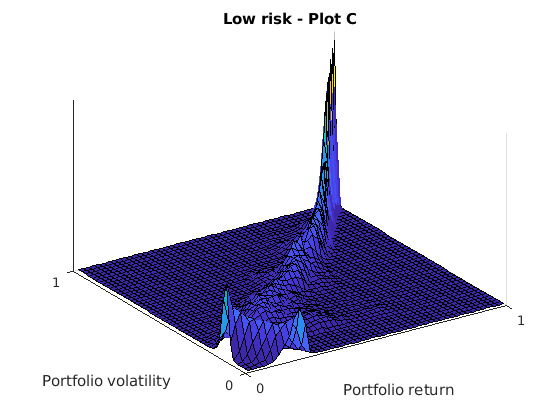}
\caption{The copulae of portfolios' return and volatility. The copula in the upper left plot (Unbiased case) is computed according to Sec.~\ref{sec:crises_detection}. We compute each one of the orthe three copulae when the portfolios have been built according to three different mixed allocation strategies, while $M=70$ strategies take place in the stock market. The mixed strategies differ based on the risk's behavioral function (high, medium, and low risk). Plot B, C, D refer to Fig.~\ref{fig:parameterization_functions}. \label{fig:copulas_strategies_w}}
\end{figure}

The upper left plot in Fig.~\ref{fig:copulas_strategies_w} illustrates the copula computed with uniform sampling from the set of long-only portfolios as in Sec.~\ref{sec:crises_detection}. The indicator of this copula is equal to $0.32$, which implies a normal relation between portfolios' return and volatility. We also obtain similar copulae when the portfolios are built according to three different mixed strategies which differ based on the risk's behavioral function (high-medium-low level of risk), as in the two previous sections and the weight vectors $w$ have been computed as in Sec.~\ref{subsubsec:param_funs}. The indicators for high, medium, and low risk are $0.002, 0.004, 0.01$ respectively.

\begin{figure}[t] \centering
\includegraphics[width=0.35\linewidth]{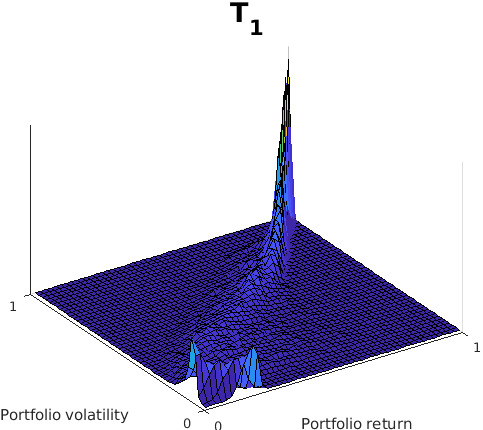}
\includegraphics[width=0.35\linewidth]{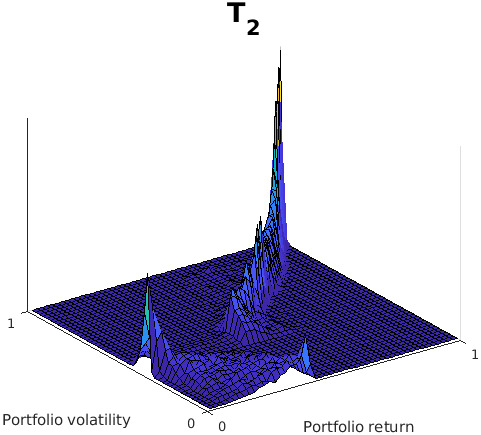}\\
\vspace{0.4cm}
\includegraphics[width=0.35\linewidth]{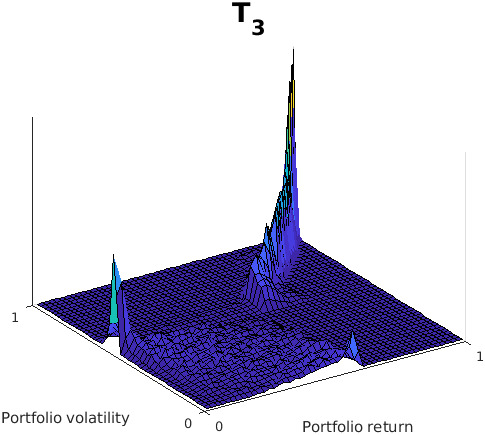}
\includegraphics[width=0.35\linewidth]{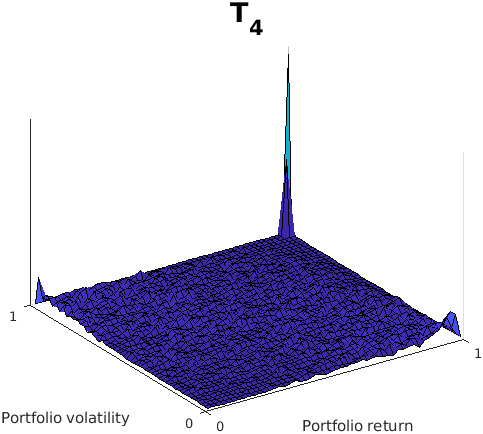}
\caption{The copulae of portfolios' return and volatility. We compute each copula for a certain mixed allocation strategy, while $M=70$ strategies take place in the stock market. We compute the bias vector using the risk's behavioral function given by Plot B in Fig.~\ref{fig:parameterization_functions}, which favors allocation strategies with a medium level of risk. We compute a copula per temperature, where $T_1>\dots >T_4$ correspond to equidistant exponential distributions $p_{T_i}$ w.r.t.\ $L_2$ norm and the temperature $T_1$ correspond to a mixed strategy with equally divided investors over the allocation strategies. \label{fig:copulas_risk_medium}}
\end{figure}

On the other hand, in Fig.~\ref{fig:copulas_risk_medium} we compute the bias vector for the risk's behavioral function given by plot B in Fig.~\ref{fig:parameterization_functions}, which favors allocation strategies with medium risk. The temperature $T_1$ corresponds to the mixed strategy with equally divided investors over the $M=70$ allocation strategies. We notice that as the temperature decreases, i.e.\ the percentage of the investors who select allocation strategies with a medium level of risk increases, a large percentage of the mass of the copula is shifting towards the corners on the down diagonal. More precisely, the indicators for $T_1>\dots >T_4$ are $0.03 ,0.03 , 0.04, 2.19$ respectively. The latter implies that the percentage of portfolios with either low return and high volatility or high return and low volatility increases as the tendency of the investors to select medium-risk allocation strategies increases in the stock market.

\section{Conclusions and future work}\label{sec:future_work}

We briefly survey existing work on crises detection and we strengthen its results employing clustering algorithms for bivariate distributions. We also use it to detect all the past crash events in the cryptocurrency markets. This problem motivates us to develop a new computational framework to model asset allocation strategies in a stock market and to define a new portfolio score based on that framework. Our simulations show that the informativeness of our new score can be higher than that of existing portfolio performance scores. To provide efficient computations we develop high dimensional MCMC samplers for log-concave distributions supported on a convex polytope. Our sampler scales up to a few hundred dimensions/assets. We simulate mixed strategies to estimate the distribution of the portfolio's score --assuming a distribution for the asset returns-- and to compute alternative copulae of portfolios' return and volatility.

A possible future work is to use nonlinear shrinkage, e.g.\ that in~\cite{Wolf20}. 
An additional direction would be to further study the state of a stock market using the alternative copula computations in Sec.~\ref{subsec:alternative_copulas}. 
Furthermore, we believe that it would be of special interest to use the distribution of the new score to define new performance measures and thus, compute the optimal portfolios with respect to those measures. In particular, the problem reduces to compute a portfolio with a ``good" distribution of score. 
Considering the copula computation in Sec.~\ref{sec:crises_detection} when the set of portfolios is the fully-invested portfolios we can not use the exact sampler in~\cite{RbMel98} because the portfolio domain $P\subset\RR^n$ is a generic convex polytope. Thus, MCMC sampling is the sole option. The latter is computationally more expensive than the method in~\cite{RbMel98}. An interesting piece of future work is to develop specialized MCMC uniform samplers for fully-invested portfolios. Last but not least, one could use the clustering methods of Sec.~\ref{sec:crises_detection} to detect intermediate states of a market.


\subsection*{Acknowledgment}
Sec.~\ref{sec:crises_detection} of this paper surveys work in collaboration with Ludovic Cal\`{e}s (JRC Ispra, Italy).

\bibliography{refs.bib}

\appendix

\section{Computational methods}\label{appnd:computational_methods}

\subsection{$L_2$ norm estimation}

To estimate the $L_2$ norm of $\pi_{\alpha_{i+1}}\propto e^{\alpha_{i+1}h(x)}$ w.r.t.\ the distribution $\pi_{\alpha_i}\propto e^{\alpha_ih(x)}$, with $\alpha_1\neq\alpha_2$ and $h(x)$ being a convex function over the set of portfolios $P$, we have,

\begin{equation}
    \begin{split}
    \| \pi_{\alpha_{i+1}} / \pi_{\alpha_i} \| & = \int_P \frac{e^{\alpha_{i+1}h(x)}}{\int_P e^{\alpha_{i+1}h(x)}dx} \frac{\int_P e^{\alpha_i h(x)}dx}{e^{\alpha_i h(x)}} \frac{e^{\alpha_{i+1}h(x)}}{\int_P e^{\alpha_{i+1}h(x)}dx} dx\\
    & = \frac{\int_P e^{\alpha_i h(x)}dx}{\int_P e^{\alpha_{i+1}h(x)}dx} \int_P e^{(\alpha_{i+1} - \alpha_i) h(x)}\frac{e^{\alpha_{i+1}h(x)}}{\int_P e^{\alpha_{i+1}h(x)}dx} dx
    \end{split}
\end{equation}

While,

\begin{equation}
    \begin{split}
    \frac{\int_P e^{\alpha_i h(x)}dx}{\int_P e^{\alpha_{i+1}h(x)}dx} & = \int_P \frac{e^{\alpha_i h(x)}}{e^{\alpha_{i+1} h(x)}}\frac{e^{\alpha_{i+1}h(x)}}{\int_P e^{\alpha_{i+1}h(x)}dx} dx\\ & = \int_P e^{(\alpha_i - \alpha_{i+1}) h(x)}\frac{e^{\alpha_{i+1}h(x)}}{\int_P e^{\alpha_{i+1}h(x)}dx} dx
    \end{split}
\end{equation}

Thus, to estimate $\| \pi_{\alpha_{i+1}} / \pi_{\alpha_i} \|$ we sample $x_1,\dots, x_k$ points from the distribution which is proportional to $e^{\alpha_{i+1} h(x)}$ and then, for sufficiently large $k$,

\begin{equation}
    \| \pi_{\alpha_{i+1}} / \pi_{\alpha_i} \| \approx \frac{\sum_{i=1}^k e^{(\alpha_i - \alpha_{i+1}) h(x_i)}}{k} \frac{\sum_{i=1}^k e^{(\alpha_{i+1} - \alpha_i) h(x_i)}}{k} 
\end{equation}

\subsection{Integral estimation}

We implement the method implied by Equation~(\ref{eq:mmc_score}). When, $\pi_i(x)\propto e^{\alpha h_i(x)}$, where $h_i(x)$ a concave function supported on $P\subset\RR^n$, we re-write the Equation~(\ref{eq:integrals}) as follows,

\begin{equation}
    s = \sum_{i=1}^Mw_i\frac{\int_{S}e^{\alpha h_i(x)}dx}{\int_{P}e^{\alpha h_i(x)}dx} .
\end{equation}
Thus, we estimate both integrals using the same method and we exactly compute $\frac{\vol(S)}{\vol(P)}$ with Varsi's algorithm~\cite{Varsi73}.

\subsection{Sampling from the set of portfolios}

Let a low-dimensional convex polytope, 

\begin{equation}
LP = \{ x\in\RR^n\ |\ Ax\leq b, A_{eq}x=b_{eq} \},\ A\in\RR^{m\times n},b\in\RR^m,A_{eq}\in\RR^{l\times n},b_{eq}\in\RR^{l} .    
\end{equation}

To compute a full dimensional polytope, we compute the matrix $N\in\RR^{n\times d}$ of the right null space of $A_{eq}$. Then, we obtain the full dimensional polytope,

\begin{equation}
    FP = \{ y\in\RR^d\ |\ By\leq z \},\ B\in\RR^{m\times d}, z\in\RR^m ,
\end{equation}
while $B = AN$ and $z = b - Ax^*$ and $x^*$ is a solution of the linear system $A_{eq}x = b_{eq}$. Moreover, the matrix $N$ defines an isometric linear transformation,
\begin{equation}
    f(y) = Ny + x^* .
\end{equation}
Thus, to sample from a log-concave distribution $\pi$ restricted to LP we transform $\pi$ according to $N$ to obtain $\pi'$. Next, we sample from $\pi'$ restricted to FP and map the generated points back to LP using the inverse linear transformation. 

To sample from a log-concave distribution truncated to a polytope we use the reflective Hamiltonian Monte Carlo given by~\cite{afshar2015reflection}.
To sample from a mixture of log-concave distribution 

\begin{equation}
    \pi(x) = \sum_i w_i\pi_i(x),
\end{equation}
we implement the following method,

\begin{enumerate}
    \item Generate $u\sim\mathcal{U}(0,1)$.
    \item If $u\in [ \sum_{i=1}^kw_i, \sum_{i=1}^{k+1}w_i ]$, generate a sample from $\pi_k$.
    \item Repeat steps 1.\ and 2.\ until you have the desired amount of samples from the mixture distribution.
\end{enumerate}

\subsection{Run-times}

In Table~\ref{tab:uniform_runtimes} we report the run-times of the exact uniform sampler in~\cite{RbMel98} and the Reflective Hamiltonian Monte Carlo we use to sample from a log-concave distribution supported on the set of portfolios. All computations were performed on a PC with {\tt Intel\textregistered\ Pentium(R) CPU G4400 @ 3.30GHz $\times$ 2 CPU} and {\tt 16GB RAM}.

\begin{table}[h!]
\centering
\begin{tabular}{c|ccccc}
 & \multicolumn{5}{c}{Number of assets / Dimension} \\
 Method    &   200    &  400    &  600    &  800 & 1000  \\
 \hline
Simpex sampler~\cite{RbMel98}  &  2.42  &  5.19  &  6.93 &    9.37 &   11.69 \\
ReHMC~\cite{afshar2015reflection} & 8.39 & 64.12 & 620.2 & -- & -- \\
\end{tabular}
\caption{The average run-times in seconds over $10$ runs for the exact uniform sampler in~\cite{RbMel98} and our MCMC method (ReHMC) to sample log-concave distributions supported on the set of portfolios $P\subset\RR^n$. For the ReHMC we sample from the spherical Gaussian centered at a uniformly distributed point in $\Delta^{n-1}$. We set an upper bound on the run-time; if the run-time exceeds $1$ hour we stop the execution.
\label{tab:uniform_runtimes}}
\end{table}

\section{Crypto data}\label{appnd:crypto_data}

 \begin{table}[h!]
\centering
\begin{tabular}{l|c|c}
Coin & Symbol & Dates\\
\hline
Bitcoin	 & BTC & 28/04/2013 - 21/11/2020\\
Litecoin & LTC & 28/04/2013 - 21/11/2020\\
Ethereum & ETH & 07/08/2015 - 21/11/2020\\
XRP & XRP & 04/08/2013 - 21/11/2020\\
Monero & XMR & 21/05/2014 - 21/11/2020\\
Tether & USDT & 25/02/2015 - 21/11/2020\\
Dash & DASH & 14/02/2014 - 21/11/2020\\
Stellar & XLM & 05/08/2014 - 21/11/2020\\
Dogecoin & DOGE & 15/12/2013 - 21/11/2020\\
DigiByte & DGB  & 06/02/2014 - 21/11/2020\\
NEM & XEM & 01/04/2015 - 21/11/2020\\
Siacoin & SC & 26/08/2015 - 21/11/2020\\
\end{tabular}
\caption{Cryptocurrencies used to detect shock events in market.}
\label{Tab:cryptos}
\end{table}

The estimation of the covariance matrix $\Sigma$ we use in Section~\ref{sec:simulations} is: 

\[
\tiny
\Sigma = 10^{-3} \left[ \begin{array}{rrrrrrrrrrrr} 
    1.4290  &  0.9808 &   0.5099  &  0.3778   & 0.6328  &  0.8382  &  0.5957  &  0.7565  &  0.4754 &   0.9317  &  1.0839  &  0.2413 \\
    0.9808  &  3.4804 &   0.7195 &   0.4652   & 1.1354 &   0.8545  &  1.1496   & 1.3633  &  0.8549   & 1.6933 &   1.3762  &  0.4291 \\
    0.5099   & 0.7195   & 4.3024  &  0.3100   & 0.9297 &   0.6226  &  0.9011   & 1.3825 &   0.4979  &  1.0408 &   0.7013 &   0.3027 \\
    0.3778  &  0.4652   & 0.3100  &  0.5014 &   0.3156  &  0.4654  &  0.3230  &  0.4687  &  0.2862  &  0.4977 &   0.4366  &  0.1317 \\
    0.6328  &  1.1354   & 0.9297   & 0.3156   & 2.5198  &  0.6002  &  1.1857   & 0.9297   & 0.5557  &  1.9265 &   0.9451  &  0.3790 \\
    0.8382  &  0.8545  &  0.6226  &  0.4654  &  0.6002  &  2.4615   & 0.5202  &  0.7768  &  0.5866   & 1.0493  &  1.1255   & 0.2875 \\
    0.5957  &  1.1496  &  0.9011   & 0.3230  &  1.1857   & 0.5202 &   3.7467  &  1.0624   & 0.5390   & 2.0051  &  1.1436   & 0.3894 \\
    0.7565 &   1.3633   & 1.3825  &  0.4687  &  0.9297   & 0.7768 &   1.0624 &   4.5856 &   0.8172 &   1.0103  &  1.0801 &   0.3444 \\
    0.4754   & 0.8549  &  0.4979  &  0.2862   & 0.5557  &  0.5866 &   0.5390   & 0.8172   & 2.2931 &   0.9156  &  0.5850 &   0.3067 \\
    0.9317   & 1.6933   & 1.0408   & 0.4977 &   1.9265  &  1.0493   & 2.0051  &  1.0103 &   0.9156 &   12.7352  &  1.2763  &  0.7029 \\
    1.0839 &   1.3762   & 0.7013 &   0.4366  &  0.9451 &   1.1255 &   1.1436  &  1.0801  &  0.5850  &  1.2763 &   4.4734   & 0.2613 \\
    0.2413   & 0.4291  &  0.3027  &  0.1317  &  0.3790   & 0.2875  &  0.3894  &  0.3444   & 0.3067  &  0.7029  &  0.2613   & 0.5760 \\
\end{array} \right] 
\]

\begin{table}[h!]
\centering
\begin{tabular}{ccccccc}
 BTC    &   LTC    &  ETH    &  XRP    &    XMR  &  USDT \\
0.44\%  &  1.10\%  &  0.45\%   -0.07\%  &  0.53  &  0.11  \\

\hline
 DASH  &   XLM    &  DOGE     &  DGB  &  XEM  &  SC  \\
0.48\%  &  0.96\% &    0.64\% &   0.96\% & 1.05\% &  -0.07\% \\

\end{tabular}
\caption{The average assets' returns from 22/10/2016 until 29/01/2017.
\label{tab:avg_asset_returns}}
\end{table}

\begin{table}[h!]
\centering
\begin{tabular}{ccccccc}
 BTC    &   LTC    &  ETH    &  XRP    &    XMR  &  USDT \\
0.10\%  &  2.93\%  & -0.02\% & 1.80\%  &  0.83\% & 4.39\%  \\

\hline
 DASH  &   XLM    &  DOGE     &  DGB  &  XEM  &  SC  \\
-2.89\% &   0.05\% &   1.70\& &   10.50\% &    0.41\% &  -0.03\% \\

\end{tabular}
\caption{The average assets' returns from 30/01/2017 until 08/02/2017.
\label{tab:avg_asset_returns}}
\end{table}

\section{Clustering of copulae}\label{appnd:clustering_copulae}

\begin{figure}[h!]
    \centering
    \includegraphics[width=.5\linewidth]{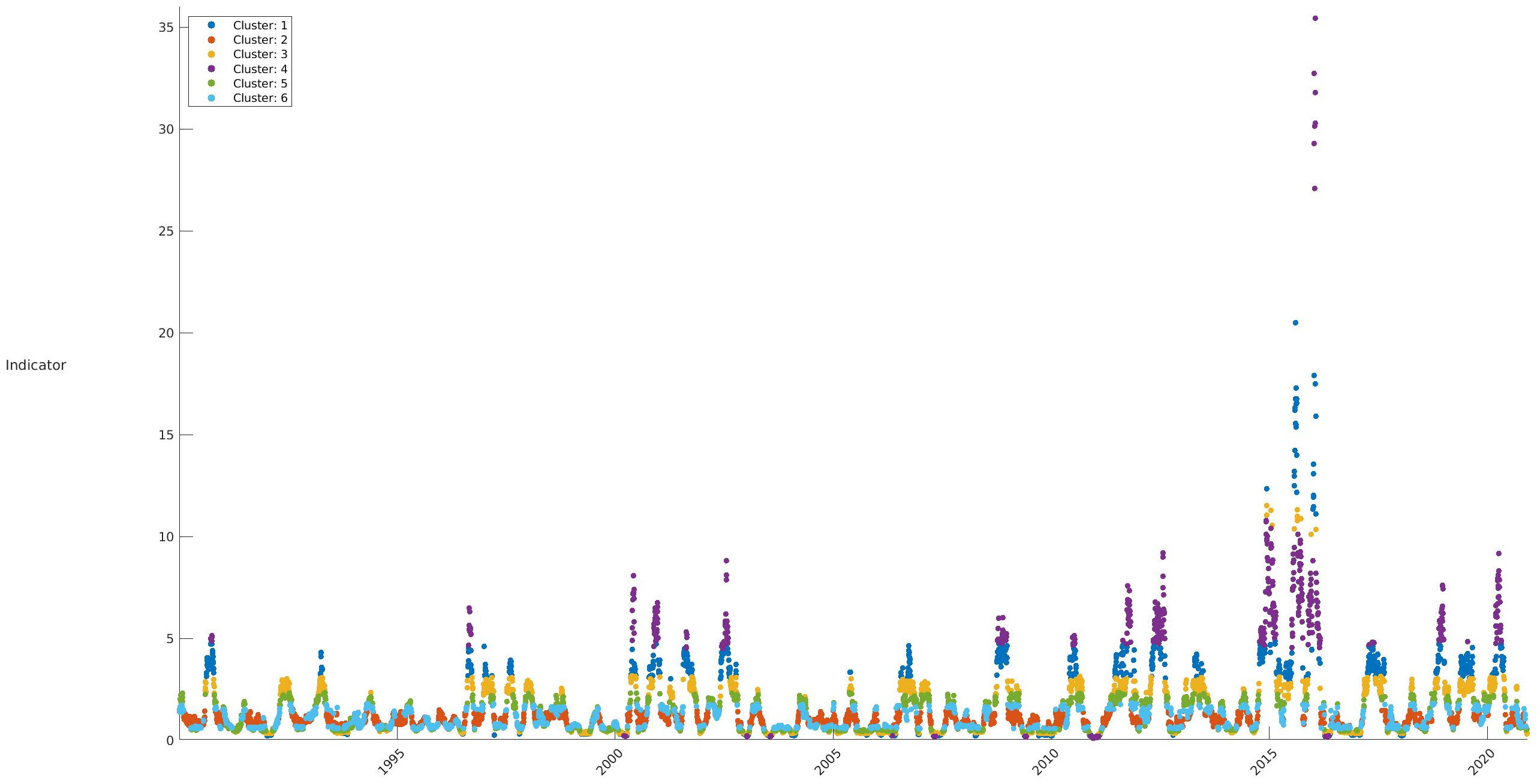}%
    \includegraphics[width=.5\linewidth]{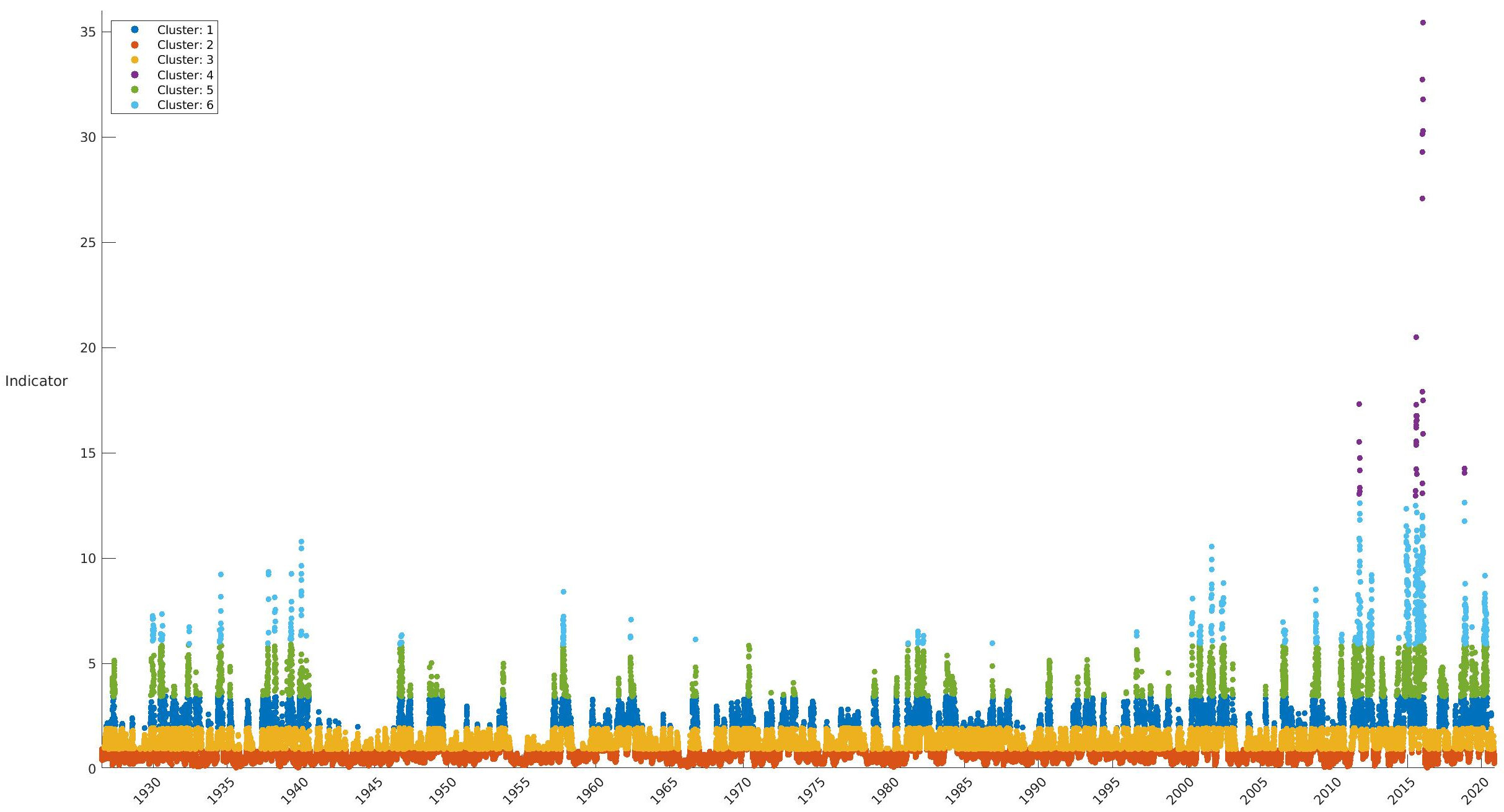}
    \caption{Left, spectral clustering ($k=6$) on EMD matrix. Right, k-medoids ($k=6$) on copulae features. Clusters appear to contain similar indicator values.}

    \label{fig:copulaeClusteringC6_clusters}
\end{figure}

\begin{figure}[h!]
    \centering
    \includegraphics[width=0.5\linewidth]{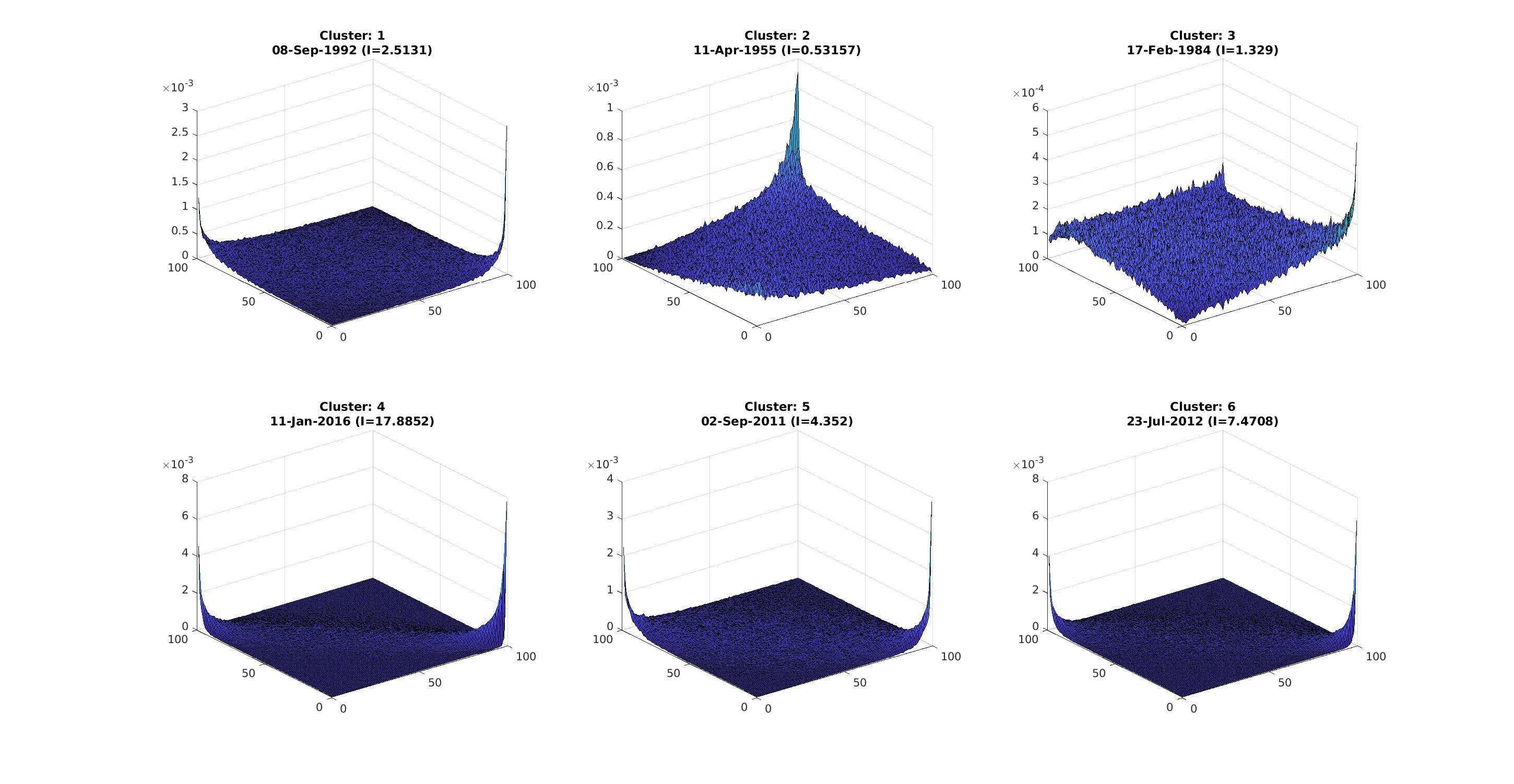}%
    \includegraphics[width=0.5\linewidth]{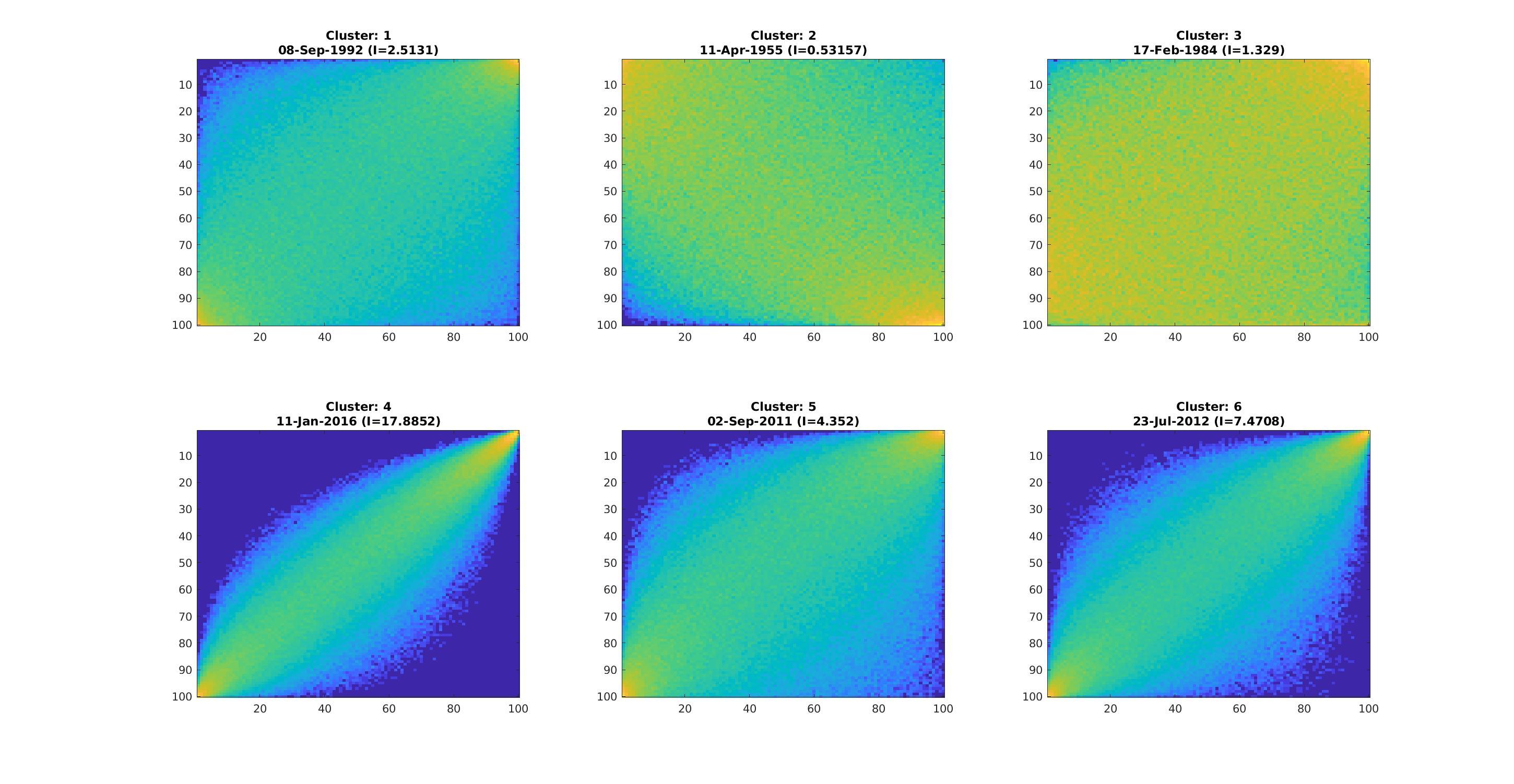}\\
    \includegraphics[width=.8\linewidth]{./clusters_C6_copulaFeats.jpeg}
    \caption{Clustering using k-medoids on copulae features.}
    \label{fig:copulaeClustering}
\end{figure}

Also, we apply clustering on features generated form the copulae, based on the indicator. We generate vector representations for each copula using the rates between all the possible combinations of the indicators’ corners: for $U_L$, $U_R$ being the upper left and right corner of a copula respectively, and for $L_L$, $L_R$ the lower left and right corners, the vector representation is  $[\frac{U_L}{U_R}\frac{U_L}{L_L}\frac{U_L}{L_R}\frac{U_R}{L_L}\frac{U_R}{L_R}\frac{L_L}{L_R}]$. These representations allow us to use clustering, such as k-medoids. Results of the clustering also follow the values of the indicator as expected (Fig. ~\ref{fig:copulaeClusteringC6_clusters}, \ref{fig:copulaeClustering}).

\begin{figure}[h!]
    \centering
    \includegraphics[width=0.5\linewidth]{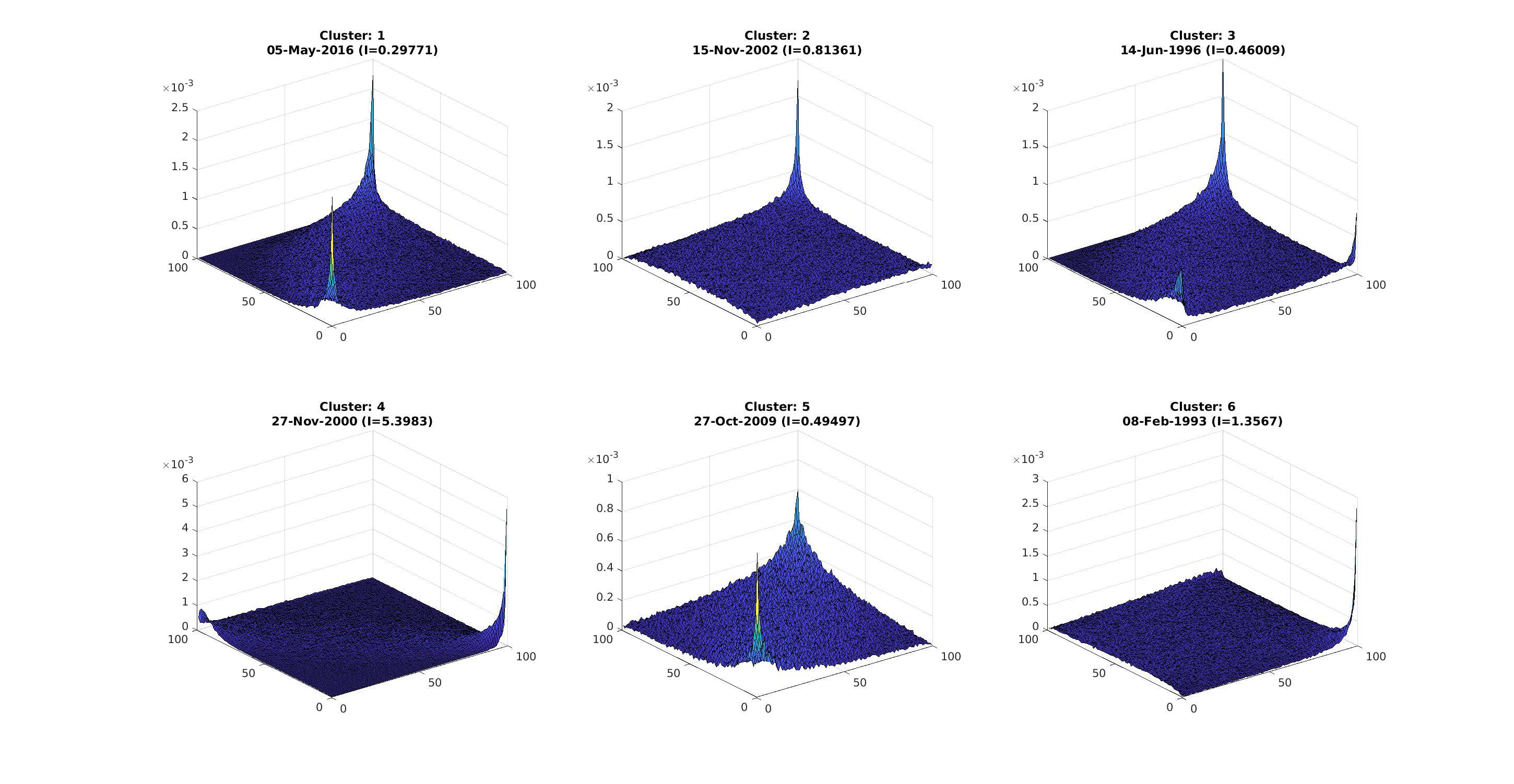}%
    \includegraphics[width=0.5\linewidth]{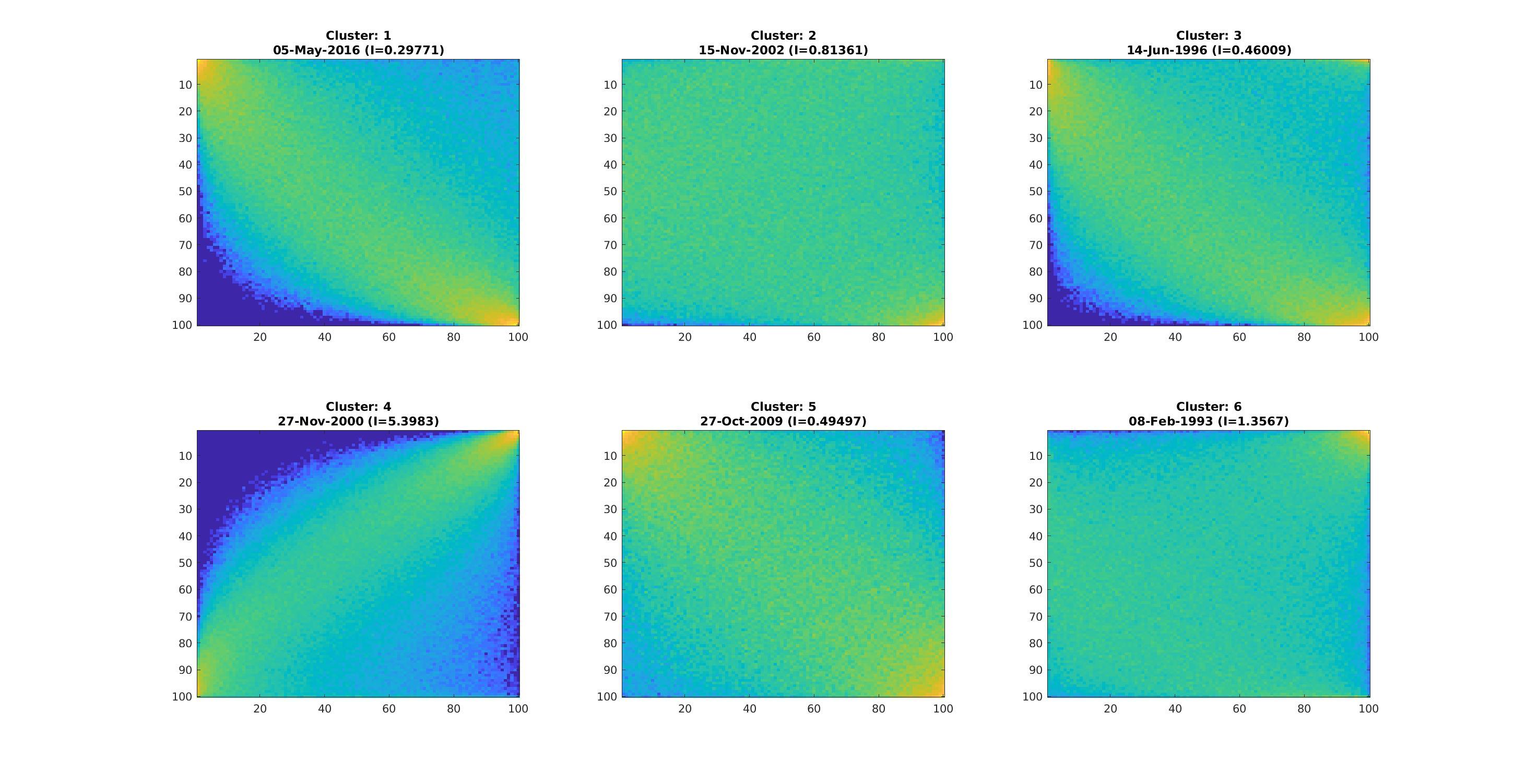}\\
    \includegraphics[width=.8\linewidth]{./clusters_C6_emd.jpeg}
    \caption{Clustering of copulae using spectral clustering on EMD distances with $k=6$.}
    \label{fig:copulaeClusteringC6EMD}
\end{figure}

\begin{figure}[h!]
    \centering
    \includegraphics[width=0.5\linewidth]{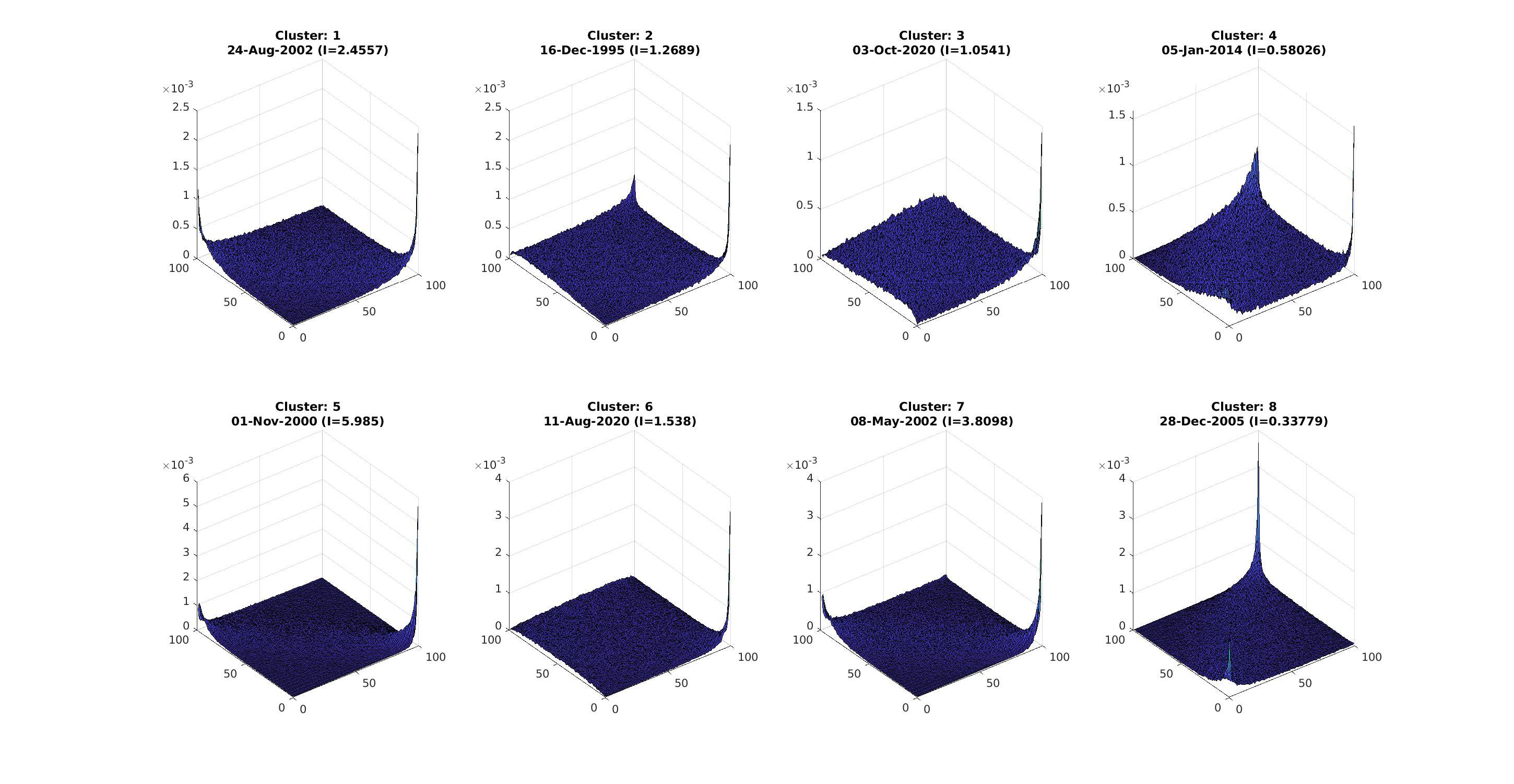}%
    \includegraphics[width=0.5\linewidth]{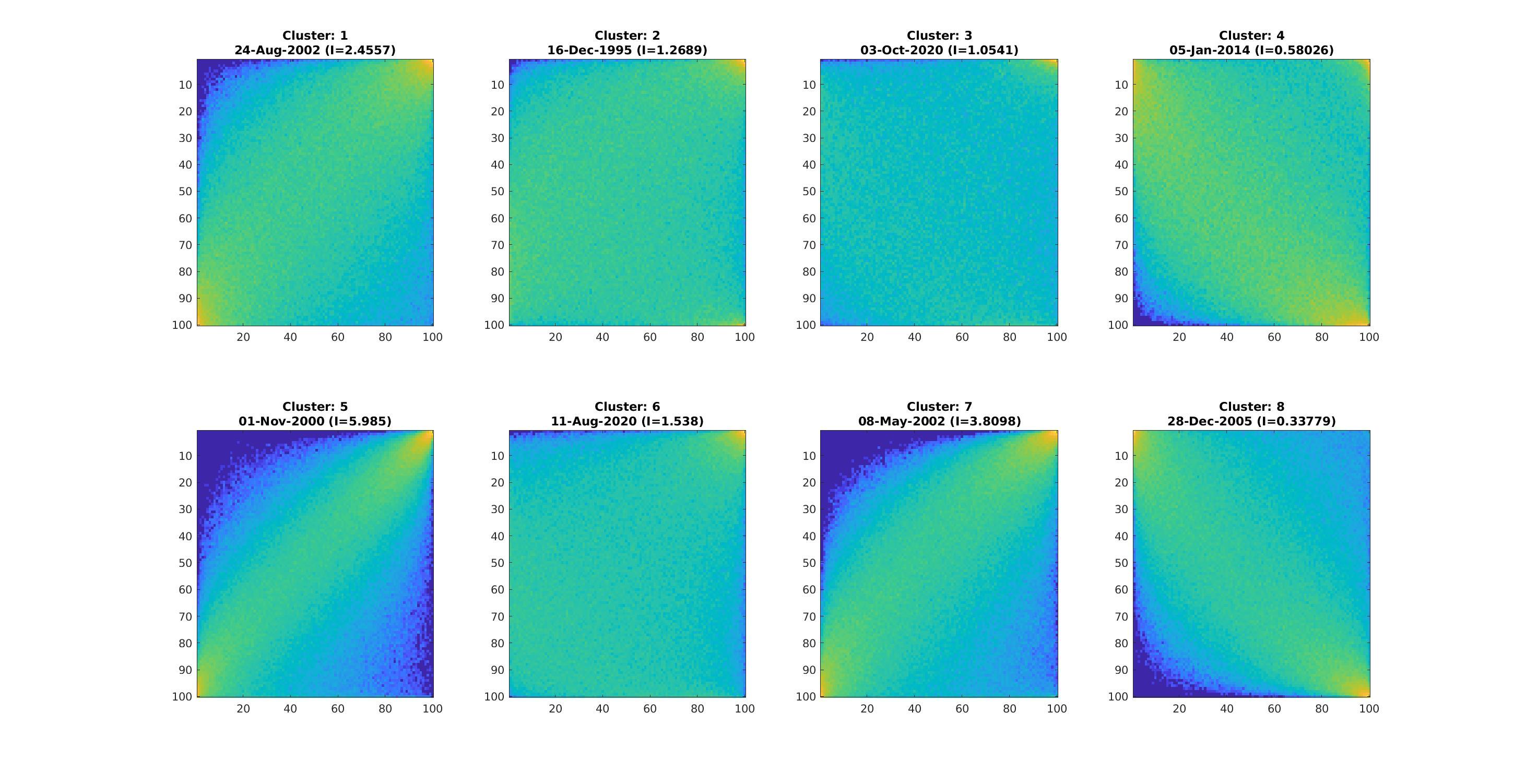}\\
    \includegraphics[width=.8\linewidth]{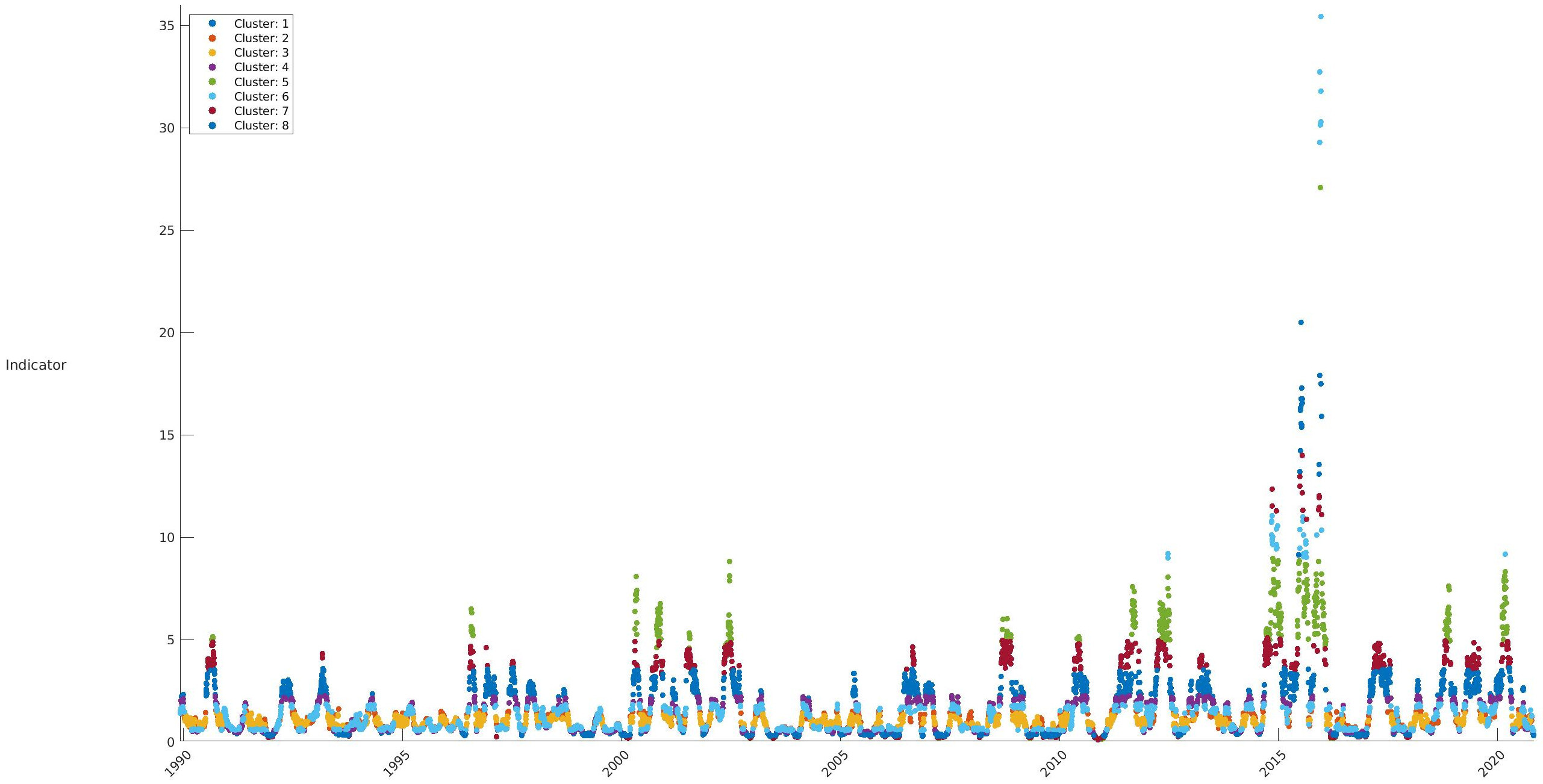}
    \caption{Clustering of copulae using spectral clustering on EMD distances with $k=8$.}
    \label{fig:copulaeClusteringC8EMD}
\end{figure}

\end{document}